\documentstyle[12pt,epsfig,amstex] {article}
\begin{document}
\def \etal  {\hbox{\it et al.}}
\def \zphys {Z. Phys.~}
\def \NIM   {Nucl. Instrum. and Methods~}
\def \PL    {Phys. Lett.~}
%
\def \gev    {{\mathrm{GeV}}}
\def \gevc   {\gev/c}
\def \gevcc  {\gev/c^2}
\def \mev    {{\mathrm{MeV}}}
\def \mevc   {\mev/c}
\def \mevcc  {\mev/c^2}
\def \mevvcc {\mev^2/c^2}
\def \dedx   {{dE/dx}}

\def \ch    {{\mathrm{c}}}
\def \b    {{\mathrm{b}}}
\def \C    {{\mathrm{C}}}
\def \B    {{\mathrm{B}}}
\def \g    {{\mathrm{g}}}
\def \q    {{\mathrm{q}}}
\def \p    {{\mathrm{p}}}
\def \D    {{\mathrm{D}}}
\def \K    {{\mathrm{K}}}
\def \Q    {{\mathrm{Q}}}
\def \Z    {{\mathrm{Z}}}
\def \uds  {{\mathrm{uds}}}
\def \udsc {{\mathrm{udsc}}}
 
\def \stat {{\mathrm{stat.}}}
\def \syst {{\mathrm{syst.}}}
 
\def \qp   {p}
\def \qpt  {p_{\perp}}
\def \ptt  {p_{\perp}^2}
\def \pt   {p_{\perp}}
%
\def \LEP    {{\sc LEP}}
\def \A      {{ALEPH}}
\def \Ab     {{{ALEPH}}}
\def \TPC    {{\sc tpc}}
\def \CLEO   {{\sc Cleo}}
\def \DELC   {{\sc Delco}}
\def \MARK   {{\sc Mark~III}}
\def \ARGUS  {{\sc Argus}} 
%
\newcommand {\goto} {\rightarrow}
\newcommand {\Prob} {{\cal{P}}(\qp,\qpt)}
\newcommand {\xc} {\langle X_{\ch} \rangle}
\newcommand {\xe} {X_E}
\newcommand {\eb} {E_{\rm{beam}}}
\newcommand {\btol} {\b\goto\ell}
\newcommand {\ctol} {\ch \to\ell}
\newcommand {\utol} {\uds\to\ell}
\newcommand {\btoctol} {\b \to \ch \to \ell}
\newcommand {\qstol} {\q\leadsto\ell}
\newcommand {\bstol} {\b\leadsto\ell}
\newcommand {\cstol} {\ch\leadsto\ell}
\newcommand {\ustol} {\uds\leadsto\ell}
%
 
\def \ec   {\epsilon_{\ch}}
\def \bb    {\b \bar{\b}}
\def \cc    {\ch \bar{\ch}}
\def \Bb    {\B \b}
\def \Bc    {\B \ch}
\def \qq    {\q \bar{\q}}
\def \QQ    {\Q \bar{\Q}}
\def \Zbb   {\Z \to \bb}
\def \Zcc   {\Z \to \cc}
\def \Zqq   {\Z \to \qq}
\def \glucc {\mathrm{g} \to \cc}
\def \glubb {\mathrm{g} \to \bb}
\def\nbar{\mbox{$\bar{n}$}}
\def\bbar{\mbox{$\bar{\mbox{b}}$} }
\def\cbar{\mbox{$\bar{\mbox{c}}$} }
\def\Qbar{\mbox{$\bar{\mbox{Q}}$} }
\def\ngcc{$\nbar_{\mbox{\scriptsize g}\rightarrow\mbox{\scriptsize c\cbar}}$}
\def\gcc{$\mbox{g}\rightarrow\mbox{c}\cbar$ }
\def\gbb{$\mbox{g}\rightarrow\mbox{b}\bbar$ }
\def\gQQ{$\mbox{g}\rightarrow\mbox{Q}\Qbar$ }
\def\ngccm{\nbar_{\mbox{\scriptsize g}\rightarrow\mbox{\scriptsize c\cbar}}}
\def\ngbbm{\nbar_{\mbox{\scriptsize g}\rightarrow\mbox{\scriptsize b\bbar}}}
\def\egcc{\epsilon_{\mbox{\scriptsize g}\rightarrow\mbox{\scriptsize c\cbar}}}
\def\egbb{\epsilon_{\mbox{\scriptsize g}\rightarrow\mbox{\scriptsize b\bbar}}}

\def \RQ    {{\mathrm{R}}_{\Q}}
\def \Rb    {{\mathrm{R}}_{\b}}
\def \Rc    {{\mathrm{R}}_{\ch}}
\def\psig   {{\cal P} (\ptt)}
\def\pqsig  {{\cal P}_\qq (\ptt)}
\def\pcsig  {{\cal P}_\cc (\ptt)}
\def\pbsig  {{\cal P}_\bb (\ptt)}
\def\pgsig  {{\cal P}_\g (\ptt)}
\def\pcsigj {{\cal P}^j_\cc (\ptt)}
\def\pbsigj {{\cal P}^j_\bb (\ptt)}
\def\pgsigj {{\cal P}^j_\g (\ptt)}
\def\pcfake {{\cal P}_\cc^{\mathrm{fake}} (\ptt)}
\def\pbfake {{\cal P}_\bb^{\mathrm{fake}} (\ptt)}

\def\kpp {\D^+\! \to \! \K^- \pi^+ \pi^+}
\def\kppp {\D^0\! \to \! \K^- \pi^+ \pi^+ \pi^-}
\def\lcpkpi {\lc\! \to \! \p \K^- \pi^+ }   
\def\kppz {\D^0\! \to \! \K^- \pi^+ \pi^0}    
\def\sz   {\D^0\! \to \! \K^- \pi^+ (\pi^0)}    
\def\kp   {\D^0\! \to \! \K^- \pi^+}    
\def\kk   {\D^0\! \to \! \K^- \K^+}
\def\KPPP {\K^- \pi^+ \pi^+ \pi^-}    
\def\KPPZ {\K^- \pi^+ \pi^0}    
\def\SZ   {\K^- \pi^+ (\pi^0)}
\def\KP   {\K \pi}
\def\dz   {\D^0}
\def\dbarounon{ ${ \hbox{D}\hbox{\kern -.38cm\lower -.28cm
\hbox{\hbox{\fivebf (---)}}} }^0$ }
\def\dm   {\Delta M}    
\def\br   {{\cal B}(\kp)}
\def\brs   {{\cal B}(\dstdec)}
\def\dst  {\D^*}
\def\dstp {\D^{*+}}
\def\dstm {\D^{*-}}
\def\dstpm {\D^{*\pm}}
\def\dpm {\D^{\pm}}
\def\dpl {\D^{+}}
\def\ds {\D_{\rm{s}}^{\pm}}
\def\dsp {\D_{\rm{s}}^+}
\def\dsstpm {\D_{\rm{s}}^{* \pm}}
\def\dsstp {\D_{\rm{s}}^{* +}}
\def\dsstm {\D_{\rm{s}}^{* -}}
\def\dsst {\D_{\rm{s}}^{*}}
\def\lc {\Lambda_{\rm{c}}^+}
\def\lcpm {\Lambda_{\rm{c}}^{+}/{\bar{\Lambda}}_{\rm{c}}^{-}}
\def\dstkp {\dstdec ,\ \kp}
\def\dstkppz {\dstdec ,\ \kppz}
\def\ps   {\pi_s}
\def\psp  {\pi_s^+}
\def\psm  {\pi_s^-}
\def\dstdec {\dstp\! \to \!\D^0 \pi^+}
\def\dstdecm {\dstm\! \to \!\bar{\D}^0 \psm}
\def\thk  {\theta_\K^*}
\def\thj  {\theta_{\rm{jet}}}
\def\ths  {\theta^*}
\def\thd  {\theta_{\D^*\!,\ \rm{jet}}}
\def\fsig {{\cal F}_{\rm{sig}}}
\def\fback {{\cal F}_{\rm{back}}}
\def\fbg  {{\cal F}_{\rm{BG}}}
\def\btoc {P_{\b \to \dst}/P_{\ch \to \dst}}
\def\pcdst {P_{\ch \to \dst}}
\def\pbdst {P_{\b \to \dst}}
\def\pqdst {P_{\q \to \dst}}
\def\epsc {\epsilon_\ch}
\def\epsb {\epsilon_\b}
\def\epsdq {\epsilon^{\q\to\KP}}
\def\epsdc {\epsilon^{\ch\to\KP}}
\def\epsdb {\epsilon^{\b\to\KP}}
\def\epspq {\epsilon^{\q\to\pi}}
\def\epspc {\epsilon^{\q\to\pi}}
\def\epspb {\epsilon^{\q\to\pi}} 
\def\epsm {\epsilon^{p_{\pi}}}
\def\epsdqrec {\epsilon_{rec}^{\q\to\KP}}
\def\epsdcrec {\epsilon_{rec}^{\ch\to\KP}}
\def\epsdbrec {\epsilon_{rec}^{\b\to\KP}}
\def\epspqrec {\epsilon_{rec}^{\q\to\pi}}
\def\epspcrec {\epsilon_{rec}^{\ch\to\pi}}
\def\epspbrec {\epsilon_{rec}^{\b\to\pi}}
\def\rhomq     {\rho_{\q}^{p_{\pi}}}
\def\rhomc     {\rho_{\ch}^{p_{\pi}}}
\def\rhomb     {\rho_{\b}^{p_{\pi}}}
\def\nhad      {N_{\rm{had}}}
\def \rightdownarrow
 {\kern.5em
 \rule[.5ex]{.15mm}{2ex}
 {\mbox{$\kern-0.1em{\rightarrow}$}}}

\begin{titlepage}
\pagestyle{empty}
%

%
\begin{center}
{\large EUROPEAN LABORATORY FOR PARTICLE PHYSICS}
\end{center}

\vskip 1.5 cm

\begin{flushright}
{\bf CERN/EP 99-094 } \\
July 8, 1999 \\
\end{flushright}

\vspace{1.5 cm}
\begin{center}
{\LARGE\bf\boldmath Study of Charm Production in Z Decays}
\end{center}
\vspace{1.cm}
\begin{center}
{\large The ALEPH collaboration${}^*$ \\}

\vspace{1cm}

\end{center}
\vspace{1.5cm}


%

\centerline{\large\bf Abstract}
\vskip 0.6 cm
The production rates of $\dstpm$, $\dsstpm$, $\dpm$, $\dz / \bar{\D}^0$, 
$\ds$, 
and $\lcpm$ in $\Z \rightarrow \cc$ decays are measured 
using the LEP I data sample recorded by the ALEPH detector.
The fractional energy spectrum of the $\dstpm$ is well described as the
sum of three contributions: charm hadronisation, b hadron decays
and gluon splitting into a pair of heavy quarks.
The probability for a c quark to hadronise into a $\dstp$ is
found to be
$f(\ch \to \dstp) = 0.233 \pm 0.010 \mathrm{(stat.)} 
\pm 0.011 \mathrm{(syst.)}$.
The average fraction of the beam energy carried by $\dstpm$ mesons
in $\Zcc$ events is measured to be 
${\langle \xe (\dstpm) \rangle }_{\cc} = 0.4878 
\pm 0.0046 \mathrm{(stat.)} \pm 0.0061 \mathrm{(syst.)} .$
The $\dstpm$ energy and the hemisphere mass
imbalance distributions are simultaneously used to measure
the fraction of hadronic Z decays in which a gluon splits to a
$\cc$ pair:
$\bar{n}_{\glucc} = (3.23 \pm 0.48  \mathrm{(stat.)} 
\pm 0.53  \mathrm{(syst.)} )\%.$
The ratio of the Vector/(Vector+Pseudoscalar) production rates in
charmed mesons is found to be $P_V = 0.595\pm0.045$. 
The fractional
decay width of the Z into $\cc$ pairs is determined from the sum of
the production rates for various weakly decaying charmed states to be
$\Rc = 0.1738 \pm 0.0047 {\rm (stat.)}
\pm 0.0116 {\rm (syst.)}.$
\vskip 2. cm
\begin{center}
(Submitted to the European Physical Journal C) \\
\end{center}

\vskip 2.5 cm


${}^*$ See next pages for the list of authors.
\end{titlepage}

\pagestyle{plain}
\setcounter{page}{1}
\setcounter{footnote}{0}


\newpage
\pagestyle{empty}
\newpage
\small
%
\newlength{\saveparskip}
\newlength{\savetextheight}
\newlength{\savetopmargin}
\newlength{\savetextwidth}
\newlength{\saveoddsidemargin}
\newlength{\savetopsep}
\setlength{\saveparskip}{\parskip}
\setlength{\savetextheight}{\textheight}
\setlength{\savetopmargin}{\topmargin}
\setlength{\savetextwidth}{\textwidth}
\setlength{\saveoddsidemargin}{\oddsidemargin}
\setlength{\savetopsep}{\topsep}
%
%
\setlength{\parskip}{0.0cm}
\setlength{\textheight}{25.0cm}
\setlength{\topmargin}{-1.5cm}
\setlength{\textwidth}{16 cm}
\setlength{\oddsidemargin}{-0.0cm}
\setlength{\topsep}{1mm}
\pretolerance=10000
\centerline{\large\bf The ALEPH Collaboration}
\footnotesize
\vspace{0.5cm}
{\raggedbottom
\begin{sloppypar}
\samepage\noindent
R.~Barate,
D.~Decamp,
P.~Ghez,
C.~Goy,
\mbox{J.-P.~Lees},
E.~Merle,
\mbox{M.-N.~Minard},
B.~Pietrzyk
\nopagebreak
\begin{center}
\parbox{15.5cm}{\sl\samepage
Laboratoire de Physique des Particules (LAPP), IN$^{2}$P$^{3}$-CNRS,
F-74019 Annecy-le-Vieux Cedex, France}
\end{center}\end{sloppypar}
\vspace{2mm}
\begin{sloppypar}
\noindent
R.~Alemany,
M.P.~Casado,
M.~Chmeissani,
J.M.~Crespo,
E.~Fernandez,
\mbox{M.~Fernandez-Bosman},
Ll.~Garrido,$^{15}$
E.~Graug\`{e}s,
A.~Juste,
M.~Martinez,
G.~Merino,
R.~Miquel,
Ll.M.~Mir,
A.~Pacheco,
I.C.~Park,
I.~Riu
\nopagebreak
\begin{center}
\parbox{15.5cm}{\sl\samepage
Institut de F\'{i}sica d'Altes Energies, Universitat Aut\`{o}noma
de Barcelona, E-08193 Bellaterra (Barcelona), Spain$^{7}$}
\end{center}\end{sloppypar}
\vspace{2mm}
\begin{sloppypar}
\noindent
A.~Colaleo,
D.~Creanza,
M.~de~Palma,
G.~Iaselli,
G.~Maggi,
M.~Maggi,
S.~Nuzzo,
A.~Ranieri,
G.~Raso,
F.~Ruggieri,
G.~Selvaggi,
L.~Silvestris,
P.~Tempesta,
A.~Tricomi,$^{3}$
G.~Zito
\nopagebreak
\begin{center}
\parbox{15.5cm}{\sl\samepage
Dipartimento di Fisica, INFN Sezione di Bari, I-70126
Bari, Italy}
\end{center}\end{sloppypar}
\vspace{2mm}
\begin{sloppypar}
\noindent
X.~Huang,
J.~Lin,
Q. Ouyang,
T.~Wang,
Y.~Xie,
R.~Xu,
S.~Xue,
J.~Zhang,
L.~Zhang,
W.~Zhao
\nopagebreak
\begin{center}
\parbox{15.5cm}{\sl\samepage
Institute of High-Energy Physics, Academia Sinica, Beijing, The People's
Republic of China$^{8}$}
\end{center}\end{sloppypar}
\vspace{2mm}
\begin{sloppypar}
\noindent
D.~Abbaneo,
U.~Becker,$^{19}$
G.~Boix,$^{6}$
M.~Cattaneo,
F.~Cerutti,
V.~Ciulli,
G.~Dissertori,
H.~Drevermann,
R.W.~Forty,
M.~Frank,
T.C.~Greening,
A.W. Halley,
J.B.~Hansen,
J.~Harvey,
P.~Janot,
B.~Jost,
I.~Lehraus,
O.~Leroy,
P.~Mato,
A.~Minten,
A.~Moutoussi,
F.~Ranjard,
L.~Rolandi,
D.~Schlatter,
M.~Schmitt,$^{20}$
O.~Schneider,$^{2}$
P.~Spagnolo,
W.~Tejessy,
F.~Teubert,
I.R.~Tomalin,
E.~Tournefier,
A.E.~Wright
\nopagebreak
\begin{center}
\parbox{15.5cm}{\sl\samepage
European Laboratory for Particle Physics (CERN), CH-1211 Geneva 23,
Switzerland}
\end{center}\end{sloppypar}
\vspace{2mm}
\begin{sloppypar}
\noindent
Z.~Ajaltouni,
F.~Badaud,
G.~Chazelle,
O.~Deschamps,
A.~Falvard,
C.~Ferdi,
P.~Gay,
C.~Guicheney,
P.~Henrard,
J.~Jousset,
B.~Michel,
S.~Monteil,
\mbox{J-C.~Montret},
D.~Pallin,
P.~Perret,
F.~Podlyski
\nopagebreak
\begin{center}
\parbox{15.5cm}{\sl\samepage
Laboratoire de Physique Corpusculaire, Universit\'e Blaise Pascal,
IN$^{2}$P$^{3}$-CNRS, Clermont-Ferrand, F-63177 Aubi\`{e}re, France}
\end{center}\end{sloppypar}
\vspace{2mm}
\begin{sloppypar}
\noindent
J.D.~Hansen,
J.R.~Hansen,
P.H.~Hansen,
B.S.~Nilsson,
B.~Rensch,
A.~W\"a\"an\"anen
\begin{center}
\parbox{15.5cm}{\sl\samepage
Niels Bohr Institute, DK-2100 Copenhagen, Denmark$^{9}$}
\end{center}\end{sloppypar}
\vspace{2mm}
\begin{sloppypar}
\noindent
G.~Daskalakis,
A.~Kyriakis,
C.~Markou,
E.~Simopoulou,
I.~Siotis,
A.~Vayaki
\nopagebreak
\begin{center}
\parbox{15.5cm}{\sl\samepage
Nuclear Research Center Demokritos (NRCD), GR-15310 Attiki, Greece}
\end{center}\end{sloppypar}
\vspace{2mm}
\begin{sloppypar}
\noindent
A.~Blondel,
G.~Bonneaud,
\mbox{J.-C.~Brient},
A.~Roug\'{e},
M.~Rumpf,
M.~Swynghedauw,
M.~Verderi,
H.~Videau
\nopagebreak
\begin{center}
\parbox{15.5cm}{\sl\samepage
Laboratoire de Physique Nucl\'eaire et des Hautes Energies, Ecole
Polytechnique, IN$^{2}$P$^{3}$-CNRS, \mbox{F-91128} Palaiseau Cedex, France}
\end{center}\end{sloppypar}
\vspace{2mm}
\begin{sloppypar}
\noindent
E.~Focardi,
G.~Parrini,
K.~Zachariadou
\nopagebreak
\begin{center}
\parbox{15.5cm}{\sl\samepage
Dipartimento di Fisica, Universit\`a di Firenze, INFN Sezione di Firenze,
I-50125 Firenze, Italy}
\end{center}\end{sloppypar}
\vspace{2mm}
\begin{sloppypar}
\noindent
R.~Cavanaugh,
M.~Corden,
C.~Georgiopoulos
\nopagebreak
\begin{center}
\parbox{15.5cm}{\sl\samepage
Supercomputer Computations Research Institute,
Florida State University,
Tallahassee, FL 32306-4052, USA $^{13,14}$}
\end{center}\end{sloppypar}
\vspace{2mm}
\begin{sloppypar}
\noindent
A.~Antonelli,
G.~Bencivenni,
G.~Bologna,$^{4}$
F.~Bossi,
P.~Campana,
G.~Capon,
V.~Chiarella,
P.~Laurelli,
G.~Mannocchi,$^{1,5}$
F.~Murtas,
G.P.~Murtas,
L.~Passalacqua,
\mbox{M.~Pepe-Altarelli}$^{1}$
\nopagebreak
\begin{center}
\parbox{15.5cm}{\sl\samepage
Laboratori Nazionali dell'INFN (LNF-INFN), I-00044 Frascati, Italy}
\end{center}\end{sloppypar}
\vspace{2mm}
\begin{sloppypar}
\noindent
L.~Curtis,
J.G.~Lynch,
P.~Negus,
V.~O'Shea,
C.~Raine,
\mbox{P.~Teixeira-Dias},
A.S.~Thompson
\nopagebreak
\begin{center}
\parbox{15.5cm}{\sl\samepage
Department of Physics and Astronomy, University of Glasgow, Glasgow G12
8QQ,United Kingdom$^{10}$}
\end{center}\end{sloppypar}
\vspace{2mm}
\begin{sloppypar}
\noindent
O.~Buchm\"uller,
S.~Dhamotharan,
C.~Geweniger,
G.~Graefe,
P.~Hanke,
G.~Hansper,
V.~Hepp,
E.E.~Kluge,
A.~Putzer,
J.~Sommer,
K.~Tittel,
S.~Werner,$^{19}$
M.~Wunsch
\nopagebreak
\begin{center}
\parbox{15.5cm}{\sl\samepage
Institut f\"ur Hochenergiephysik, Universit\"at Heidelberg, D-69120
Heidelberg, Germany$^{16}$}
\end{center}\end{sloppypar}
\vspace{2mm}
\begin{sloppypar}
\noindent
R.~Beuselinck,
D.M.~Binnie,
W.~Cameron,
P.J.~Dornan,$^{1}$
M.~Girone,
S.~Goodsir,
E.B.~Martin,
N.~Marinelli,
A.~Sciab\`a,
J.K.~Sedgbeer,
E.~Thomson,
M.D.~Williams
\nopagebreak
\begin{center}
\parbox{15.5cm}{\sl\samepage
Department of Physics, Imperial College, London SW7 2BZ,
United Kingdom$^{10}$}
\end{center}\end{sloppypar}
\vspace{2mm}
\begin{sloppypar}
\noindent
V.M.~Ghete,
P.~Girtler,
E.~Kneringer,
D.~Kuhn,
G.~Rudolph
\nopagebreak
\begin{center}
\parbox{15.5cm}{\sl\samepage
Institut f\"ur Experimentalphysik, Universit\"at Innsbruck, A-6020
Innsbruck, Austria$^{18}$}
\end{center}\end{sloppypar}
\vspace{2mm}
\begin{sloppypar}
\noindent
C.K.~Bowdery,
P.G.~Buck,
A.J.~Finch,
F.~Foster,
G.~Hughes,
R.W.L.~Jones,
N.A.~Robertson,
\linebreak
M.I.~Williams
\nopagebreak
\begin{center}
\parbox{15.5cm}{\sl\samepage
Department of Physics, University of Lancaster, Lancaster LA1 4YB,
United Kingdom$^{10}$}
\end{center}\end{sloppypar}
\vspace{2mm}
\begin{sloppypar}
\noindent
I.~Giehl,
K.~Jakobs,
K.~Kleinknecht,
G.~Quast,
B.~Renk,
E.~Rohne,
\mbox{H.-G.~Sander},
H.~Wachsmuth,
C.~Zeitnitz
\nopagebreak
\begin{center}
\parbox{15.5cm}{\sl\samepage
Institut f\"ur Physik, Universit\"at Mainz, D-55099 Mainz, Germany$^{16}$}
\end{center}\end{sloppypar}
\vspace{2mm}
\begin{sloppypar}
\noindent
J.J.~Aubert,
C.~Benchouk,
A.~Bonissent,
J.~Carr,$^{1}$
P.~Coyle,
F.~Etienne,
F.~Motsch,
P.~Payre,
D.~Rousseau,
M.~Talby,
M.~Thulasidas
\nopagebreak
\begin{center}
\parbox{15.5cm}{\sl\samepage
Centre de Physique des Particules, Facult\'e des Sciences de Luminy,
IN$^{2}$P$^{3}$-CNRS, F-13288 Marseille, France}
\end{center}\end{sloppypar}
\vspace{2mm}
\begin{sloppypar}
\noindent
M.~Aleppo,
M.~Antonelli,
F.~Ragusa
\nopagebreak
\begin{center}
\parbox{15.5cm}{\sl\samepage
Dipartimento di Fisica, Universit\`a di Milano e INFN Sezione di Milano,
I-20133 Milano, Italy}
\end{center}\end{sloppypar}
\vspace{2mm}
\begin{sloppypar}
\noindent
V.~B\"uscher,
H.~Dietl,
G.~Ganis,
K.~H\"uttmann,
G.~L\"utjens,
C.~Mannert,
W.~M\"anner,
\mbox{H.-G.~Moser},
S.~Schael,
R.~Settles,
H.~Seywerd,
H.~Stenzel,
W.~Wiedenmann,
G.~Wolf
\nopagebreak
\begin{center}
\parbox{15.5cm}{\sl\samepage
Max-Planck-Institut f\"ur Physik, Werner-Heisenberg-Institut,
D-80805 M\"unchen, Germany\footnotemark[16]}
\end{center}\end{sloppypar}
\vspace{2mm}
\begin{sloppypar}
\noindent
P.~Azzurri,
J.~Boucrot,
O.~Callot,
S.~Chen,
A.~Cordier,
M.~Davier,
L.~Duflot,
\mbox{J.-F.~Grivaz},
Ph.~Heusse,
A.~Jacholkowska,$^{1}$
F.~Le~Diberder,
J.~Lefran\c{c}ois,
\mbox{A.-M.~Lutz},
\mbox{M.-H.~Schune},
\mbox{J.-J.~Veillet},
I.~Videau,$^{1}$
D.~Zerwas
\nopagebreak
\begin{center}
\parbox{15.5cm}{\sl\samepage
Laboratoire de l'Acc\'el\'erateur Lin\'eaire, Universit\'e de Paris-Sud,
IN$^{2}$P$^{3}$-CNRS, F-91898 Orsay Cedex, France}
\end{center}\end{sloppypar}
\vspace{2mm}
\begin{sloppypar}
\noindent
G.~Bagliesi,
S.~Bettarini,
T.~Boccali,
C.~Bozzi,$^{12}$
G.~Calderini,
R.~Dell'Orso,
I.~Ferrante,
L.~Fo\`{a},
A.~Giassi,
A.~Gregorio,
F.~Ligabue,
A.~Lusiani,
P.S.~Marrocchesi,
A.~Messineo,
F.~Palla,
G.~Rizzo,
G.~Sanguinetti,
G.~Sguazzoni,
R.~Tenchini,
C.~Vannini,
A.~Venturi,
P.G.~Verdini
\samepage
\begin{center}
\parbox{15.5cm}{\sl\samepage
Dipartimento di Fisica dell'Universit\`a, INFN Sezione di Pisa,
e Scuola Normale Superiore, I-56010 Pisa, Italy}
\end{center}\end{sloppypar}
\vspace{2mm}
\begin{sloppypar}
\noindent
G.A.~Blair,
G.~Cowan,
M.G.~Green,
T.~Medcalf,
J.A.~Strong,
\mbox{J.H.~von~Wimmersperg-Toeller}
\nopagebreak
\begin{center}
\parbox{15.5cm}{\sl\samepage
Department of Physics, Royal Holloway \& Bedford New College,
University of London, Surrey TW20 OEX, United Kingdom$^{10}$}
\end{center}\end{sloppypar}
\vspace{2mm}
\begin{sloppypar}
\noindent
D.R.~Botterill,
R.W.~Clifft,
T.R.~Edgecock,
P.R.~Norton,
J.C.~Thompson
\nopagebreak
\begin{center}
\parbox{15.5cm}{\sl\samepage
Particle Physics Dept., Rutherford Appleton Laboratory,
Chilton, Didcot, Oxon OX11 OQX, United Kingdom$^{10}$}
\end{center}\end{sloppypar}
\vspace{2mm}
\begin{sloppypar}
\noindent
\mbox{B.~Bloch-Devaux},
P.~Colas,
S.~Emery,
W.~Kozanecki,
E.~Lan\c{c}on,
\mbox{M.-C.~Lemaire},
E.~Locci,
P.~Perez,
J.~Rander,
\mbox{J.-F.~Renardy},
A.~Roussarie,
\mbox{J.-P.~Schuller},
J.~Schwindling,
A.~Trabelsi,$^{21}$
B.~Vallage
\nopagebreak
\begin{center}
\parbox{15.5cm}{\sl\samepage
CEA, DAPNIA/Service de Physique des Particules,
CE-Saclay, F-91191 Gif-sur-Yvette Cedex, France$^{17}$}
\end{center}\end{sloppypar}
\vspace{2mm}
\begin{sloppypar}
\noindent
S.N.~Black,
J.H.~Dann,
R.P.~Johnson,
H.Y.~Kim,
N.~Konstantinidis,
A.M.~Litke,
M.A. McNeil,
\linebreak
G.~Taylor
\nopagebreak
\begin{center}
\parbox{15.5cm}{\sl\samepage
Institute for Particle Physics, University of California at
Santa Cruz, Santa Cruz, CA 95064, USA$^{22}$}
\end{center}\end{sloppypar}
\vspace{2mm}
\begin{sloppypar}
\noindent
C.N.~Booth,
S.~Cartwright,
F.~Combley,
M.S.~Kelly,
M.~Lehto,
L.F.~Thompson
\nopagebreak
\begin{center}
\parbox{15.5cm}{\sl\samepage
Department of Physics, University of Sheffield, Sheffield S3 7RH,
United Kingdom$^{10}$}
\end{center}\end{sloppypar}
\vspace{2mm}
\begin{sloppypar}
\noindent
K.~Affholderbach,
A.~B\"ohrer,
S.~Brandt,
C.~Grupen,
J.~Hess,
C.~Koob,
A.~Misiejuk,
G.~Prange,
U.~Sieler
\nopagebreak
\begin{center}
\parbox{15.5cm}{\sl\samepage
Fachbereich Physik, Universit\"at Siegen, D-57068 Siegen,
 Germany$^{16}$}
\end{center}\end{sloppypar}
\vspace{2mm}
\begin{sloppypar}
\noindent
G.~Giannini,
B.~Gobbo
\nopagebreak
\begin{center}
\parbox{15.5cm}{\sl\samepage
Dipartimento di Fisica, Universit\`a di Trieste e INFN Sezione di Trieste,
I-34127 Trieste, Italy}
\end{center}\end{sloppypar}
\vspace{2mm}
\begin{sloppypar}
\noindent
J.~Rothberg,
S.~Wasserbaech
\nopagebreak
\begin{center}
\parbox{15.5cm}{\sl\samepage
Experimental Elementary Particle Physics, University of Washington, WA 98195
Seattle, U.S.A.}
\end{center}\end{sloppypar}
\vspace{2mm}
\begin{sloppypar}
\noindent
S.R.~Armstrong,
P.~Elmer,
D.P.S.~Ferguson,
Y.~Gao,
S.~Gonz\'{a}lez,
O.J.~Hayes,
H.~Hu,
S.~Jin,
P.A.~McNamara III,
J.~Nielsen,
W.~Orejudos,
Y.B.~Pan,
Y.~Saadi,
I.J.~Scott,
J.~Walsh,
Sau~Lan~Wu,
X.~Wu,
G.~Zobernig
\nopagebreak
\begin{center}
\parbox{15.5cm}{\sl\samepage
Department of Physics, University of Wisconsin, Madison, WI 53706,
USA$^{11}$}
\end{center}\end{sloppypar}
}
\footnotetext[1]{Also at CERN, 1211 Geneva 23, Switzerland.}
\footnotetext[2]{Now at Universit\'e de Lausanne, 1015 Lausanne, Switzerland.}
\footnotetext[3]{Also at Centro Siciliano di Fisica Nucleare e Struttura
della Materia, INFN, Sezione di Catania, 95129 Catania, Italy.}
\footnotetext[4]{Also Istituto di Fisica Generale, Universit\`{a} di
Torino, 10125 Torino, Italy.}
\footnotetext[5]{Also Istituto di Cosmo-Geofisica del C.N.R., Torino,
Italy.}
\footnotetext[6]{Supported by the Commission of the European Communities,
contract ERBFMBICT982894.}
\footnotetext[7]{Supported by CICYT, Spain.}
\footnotetext[8]{Supported by the National Science Foundation of China.}
\footnotetext[9]{Supported by the Danish Natural Science Research Council.}
\footnotetext[10]{Supported by the UK Particle Physics and Astronomy Research
Council.}
\footnotetext[11]{Supported by the US Department of Energy, grant
DE-FG0295-ER40896.}
\footnotetext[12]{Now at INFN Sezione de Ferrara, 44100 Ferrara, Italy.}
\footnotetext[13]{Supported by the US Department of Energy, contract
DE-FG05-92ER40742.}
\footnotetext[14]{Supported by the US Department of Energy, contract
DE-FC05-85ER250000.}
\footnotetext[15]{Permanent address: Universitat de Barcelona, 08208 Barcelona,
Spain.}
\footnotetext[16]{Supported by the Bundesministerium f\"ur Bildung,
Wissenschaft, Forschung und Technologie, Germany.}
\footnotetext[17]{Supported by the Direction des Sciences de la
Mati\`ere, C.E.A.}
\footnotetext[18]{Supported by Fonds zur F\"orderung der wissenschaftlichen
Forschung, Austria.}
\footnotetext[19]{Now at SAP AG, 69185 Walldorf, Germany.}
\footnotetext[20]{Now at Harvard University, Cambridge, MA 02138, U.S.A.}
\footnotetext[21]{Now at D\'epartement de Physique, Facult\'e des Sciences de Tunis, 1060 Le Belv\'ed\`ere, Tunisia.}
\footnotetext[22]{Supported by the US Department of Energy,
grant DE-FG03-92ER40689.}
%
%
\setlength{\parskip}{\saveparskip}
\setlength{\textheight}{\savetextheight}
\setlength{\topmargin}{\savetopmargin}
\setlength{\textwidth}{\savetextwidth}
\setlength{\oddsidemargin}{\saveoddsidemargin}
\setlength{\topsep}{\savetopsep}
\normalsize
\newpage
\pagestyle{plain}
\setcounter{page}{1}

\newpage
\section{Introduction}
Charm quarks are produced in 40\% of the hadronic $\Z$ decays,
mostly coming from $\Zcc$ or $\bb$, but also from gluon splitting
to $\cc$ or $\bb$.
This paper presents a wide-ranging survey of the production of the
weakly decaying charm states
$\dz$, $\dpl$, $\dsp$ mesons and $\lc$ baryons and
the first excited states $\dstp$, $\dsstp$ in $\Z$ decays (here and
throughout this paper, charge conjugation is implied).
It extends the work of previous publications \cite{dspap, dsubs}
to the full data sample collected by ALEPH between 1991 and 1995,
which consists of four million hadronic $\Z$ decays.
A new algorithm based on the mass of 
particles from secondary vertices
greatly improves the separation
of charm states resulting from $\bb$ and $\cc$ production.

The paper is organised as follows. After a brief description of 
relevant details of the ALEPH detector in Section 2, the procedures 
adopted to reconstruct the above charm states are given in 
Section 3. The cleanest charm signal is obtained for the $\dstp$
mesons and Section 4 is devoted to the fractional energy spectrum for their
production, which now covers the whole range from 0.1 to 1.0,
and the relative strengths of the $\bb$, $\cc$ and gluon splitting
components. In Section 5 a new technique involving the heavy and
light hemisphere masses is introduced in order to obtain a more
accurate measure of the rate for gluon splitting to charm quarks.
A measurement of the $\dsstpm$ production rate is given in Section 6.
Determinations of the production rates for the
weakly decaying charm states are presented in Section 7 and this 
information is 
used in Section 8 to yield a measurement of
$\Rc = \Gamma(\Zcc)/\Gamma(\Z \to hadrons)$ from charm counting.

\section{The \Ab\ detector}

A detailed description of the \A\ detector and its performance 
can be found in Refs.~\cite{Apparatus, Performance}.
Only a brief review is given here.

Charged particles are detected in the central part, consisting of
a two-layer silicon vertex detector with double-sided ($r$-$\phi$ and $z$)
readout, a cylindrical drift chamber and a large time projection
chamber (\TPC), which together measure up to 33 coordinates along the
charged particle trajectories.
Tracking is performed in a 1.5~T magnetic field provided by a
superconducting solenoid. 
For high momentum tracks the combined system yields a $1/p_T$
resolution of $6 \times 10^{-4} (\gevc)^{-1}$ and an impact
parameter resolution of 25~$\mu$m in both the $r$-$\phi$ and $z$
projections.
The \TPC\ also provides up to 338 measurements of ionization 
($\dedx$) allowing particle identification to be performed.
For tracks having at least 50 energy deposition 
measurements, 
the variable $\chi_h$ is defined as $(I_{\rm{meas}} - I_h)
/ \sigma_h$, where $I_{\rm{meas}}$ is the measured value of $\dedx$, 
$I_h$ is the expected value for
particle type $h$ ($h= \mathrm{p,K}$ or $\pi$), and $\sigma_h$ is the 
expected uncertainty. 

Data from the vertex detector is particularly important for 
b tagging for which a lifetime-mass
algorithm has been designed to provide discrimination
between b quarks and lighter quarks. This algorithm 
uses the significance $\cal{S}$ of the three-dimensional impact parameter 
for each charged track to define the confidence level for it
to be consistent with coming from the primary vertex. These confidence
levels are combined to obtain the confidence level $\cal{P}_{\cal{H}}$
for all tracks of an hemisphere to come from the primary vertex. The
distribution of $\cal{P}_{\cal{H}}$ is strongly peaked near zero for
b hemispheres. A second b tagging variable makes use of the b/c hadron
mass difference: the tracks in a hemisphere are grouped together
in order of decreasing
$\cal{S}$ until an effective mass of 1.8~$\gevcc$ is reached for the 
system, and the confidence level $\mu_{\cal{H}}$
for the last added track to come
from the primary vertex is used as the tagging variable. This
is peaked at zero for
b quarks. These two tagging variables are combined in a single one:
${{\cal{B}}_{tag}} = -(0.7 log_{10}  \mu_{\cal{H}} + 
0.3 log_{10} {\cal{P}}_{\cal{H}}.$
A detailed description of this algorithm can be found in Ref.~\cite{qmbtag}. 

The electromagnetic calorimeter is a lead/wire-chamber sandwich
operated in proportional mode. It is read out in projective towers
of typically $15 \times 15$~mrad$^2$ size segmented in three longitudinal
sections.
The iron return yoke is instrumented with streamer tubes
read out in projective towers and this provides the hadron calorimetry.

The tracking and calorimetry information is combined in an energy flow
algorithm providing a list of objects classified as tracks, photons 
and neutral hadrons. Particle tracks which originate from within a
cylinder of 2 cm radius and 20 cm length centred on the nominal 
interaction point, and which have more than 4 TPC hits,
are defined as {\it good tracks}. Hadronic Z decays are selected by
requiring the presence of at least 5 such good tracks and a total
visible energy greater than $10 \%$ of the centre-of-mass 
energy~\cite{hadsel}.

\section{\label{cansel}Reconstruction of Charmed Hadron Decays}

Each charm hadron species is reconstructed in the decay mode best suited to
a rate measurement, i.e. with a large enough and accurately known
branching ratio, and a sufficiently low background.

\begin{figure}[htbp]\centering
\epsfxsize=250pt
\mbox{\epsfig{file=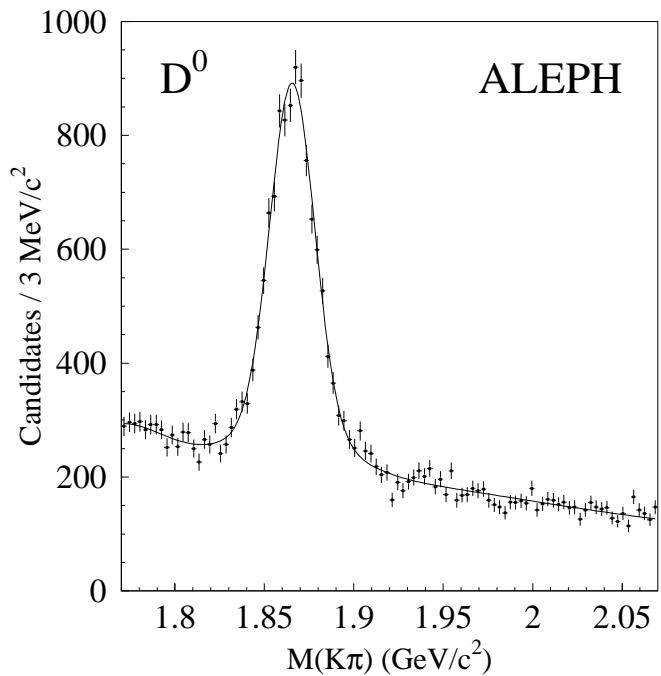,height=280 pt}}
\caption{{\small $\K^-\pi^+$ invariant mass distribution. The fitted curve is
described in section 7.}}
\label{d0}
\end{figure}

\boldmath
\subsection{$\dz$ Meson}
\unboldmath
$\dz$ mesons are selected through their decay mode $\kp$. 
A track with a momentum greater than $2.5~\gevc$, 
to which the kaon mass is assigned,
is combined with a track of opposite charge with a momentum greater than 
$1.5~\gevc$, which is assumed to be a pion. 
The contribution due to wrong mass assignment of the two tracks from the $\dz$
is reduced by means of the $\dedx$ measurement of the kaon candidate when 
available.
The $\dedx$ of the kaon track must be closer to the expectation for
a kaon than that for a pion: $|\chi_{\K}|<|\chi_{\pi}|$.
 The pair formed by the two tracks is retained if its fractional energy   
$\xe = E/\eb$ is greater than 0.5 and if a common vertex with a 
$\chi^2$ probability greater than $1\%$ is found.
The significance of the $\dz$ decay length, projected on the $\dz$ momentum,
is required to be greater than 1. 
The invariant mass distribution is shown 
in Fig.~\ref{d0}.

\begin{figure}[htbp]\centering
  \setlength{\unitlength}{1.0mm}
  \begin{picture}(150,150)
    \put(-12,-50){
    \mbox{\epsfig{file=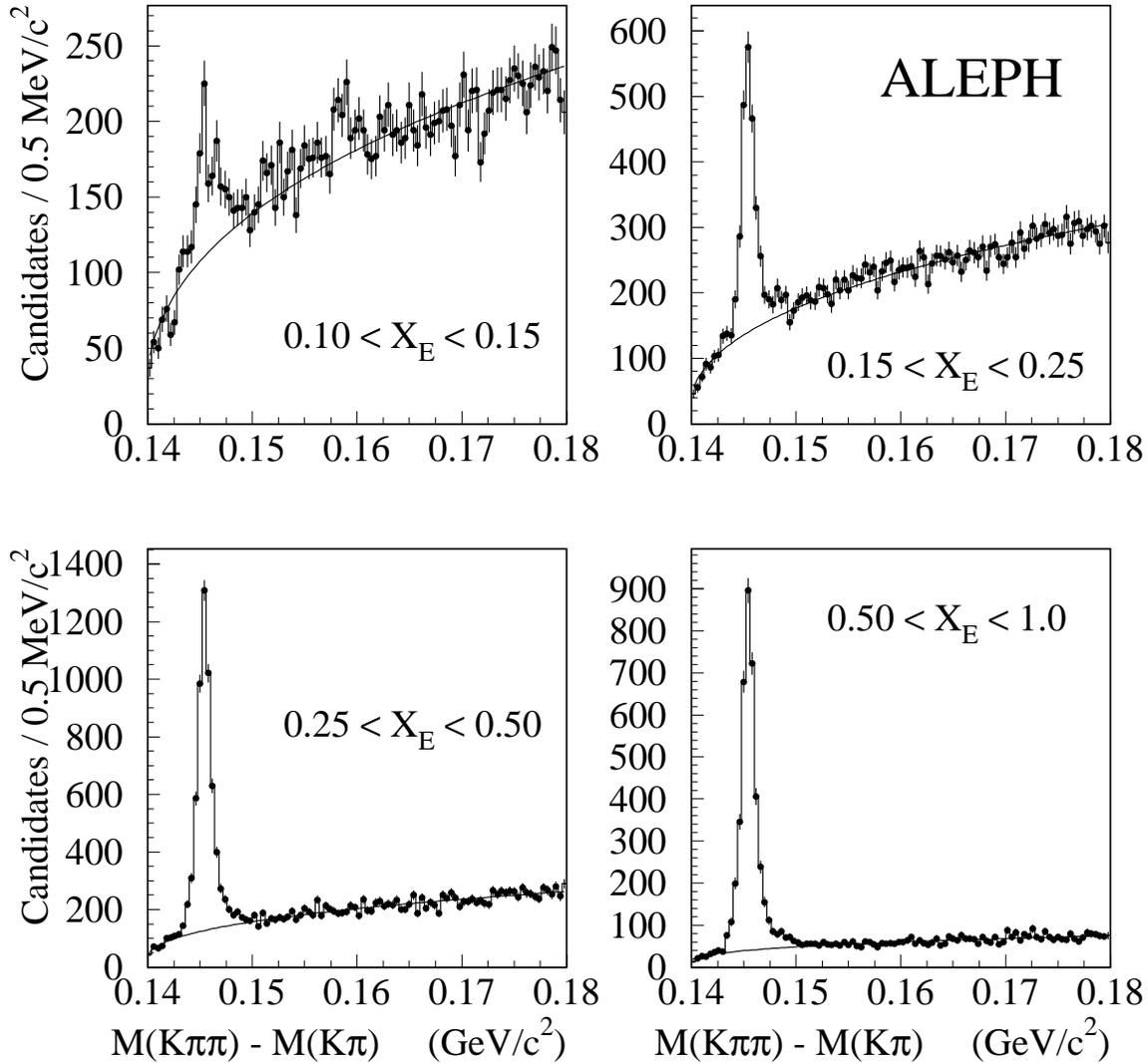,height=17cm}}
    }
  \end{picture}
    \caption{{\small Distribution of the $\K \pi \pi$ - $\K \pi$ mass
 difference in four different ranges of the $\K \pi \pi$ energy,
after requiring that the $\K\pi$ mass is consistent with the $\dz$ mass
(histogram with error bars).
The background fit is also shown.}}
    \label{deltam}
\end{figure}

\boldmath
\subsection{$\dstp$ Meson} 
\unboldmath

The $\dstkp$ decay chain is reconstructed by considering triplets of charged 
tracks with total charge $\pm 1$. 
The pion mass is assigned to the two particles with 
the same charge, and the kaon mass 
to the remaining one. 
The total energy of the triplet is required to be greater than 
$0.1 \times \eb$. 
Any $\K^{-} \pi^{+}$ combination whose mass
is within $30~\mevcc$ of the $\dz$ mass is kept.
For such a combination, the mass difference $M(\K \pi \pi) - M(\K \pi)$
is calculated and required to be within $1.6~\mevcc$ of the 
nominal $\dstp-\dz$
mass difference.
The combinatorial background level is a steeply falling function of $\xe$
and background rejection must therefore be enhanced as one goes to 
lower $\xe$. 
For this, the scalar nature of the $\dz$ and its significant
lifetime are used as follows. The distribution
of the cosine of the $\dz$ decay angle (defined in the $\K\pi$ rest frame
as the angle $\theta^*$ of the kaon
with the $\K\pi$ line of flight in the laboratory) 
is isotropic, whereas the background peaks forward and backward,
and so, for $\xe < 0.45$ it is required that $ -0.8 < \cos \theta^* < 0.9$.
The decay length significance is required to be positive for 
$0.3 < \xe < 0.4$ and greater than 1 for $\xe$ less than 0.3. 
The $\chi^2$ probability for the secondary vertex fit is
required to be greater than $1 \%$ for $\xe < 0.15$. 
The mass-difference distributions in four $\xe$ ranges are 
shown in Fig.~\ref{deltam}.

 
\boldmath
\subsection{$\dpl$ Meson}
\begin{figure}[htbp]\centering
\mbox{\epsfig{file=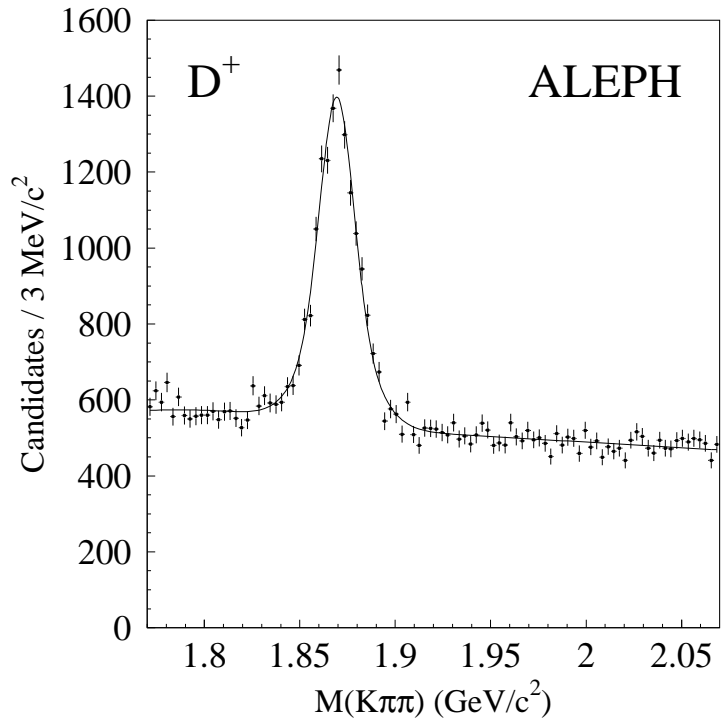,height=10.5cm}}
\caption{{\small $\K^-\pi^+\pi^+$ invariant mass distribution 
with the fit result.}}
\label{dp}
\end{figure}

\unboldmath 
The $\dpl$ meson is reconstructed in the decay mode $\kpp$. Combinations of 
three tracks are formed with the same mass assignment as for the 
$\dstp$ reconstruction. To fight the high background of this 
decay, 
the kaon track candidate is required to have
a momentum greater than $2.5~\gevc$ and an associated $\dedx$
satisfying $\chi_{\K}+\chi_{\pi} <1$, 
one of the two pion tracks must have a momentum greater than $1.5~\gevc$, 
while the other must have more than $0.75~\gevc$.
Both of the possible $\K\pi$ combinations
are required to have a $\K\pi\pi$-$\K\pi$
mass difference larger than $0.15~\gevcc$, in order to reject 
candidates consistent with the decay $\D^{*+}\to \pi^+ \dz \to \pi^+
\K^-\pi^+ X$.
Triplets of tracks with a total energy greater than $0.5 \times \eb$ and 
forming a common vertex with a $\chi^2$ probability greater than $1\%$ 
are retained.
A projected decay length significance greater than 1.5 is required
for the reconstructed $\dpl$ vertices.
The invariant mass 
distribution is shown in Fig.~\ref{dp}.

\boldmath
\subsection{$\dsp$ Meson}
\begin{figure}[htbp]\centering
    \mbox{\epsfig{file=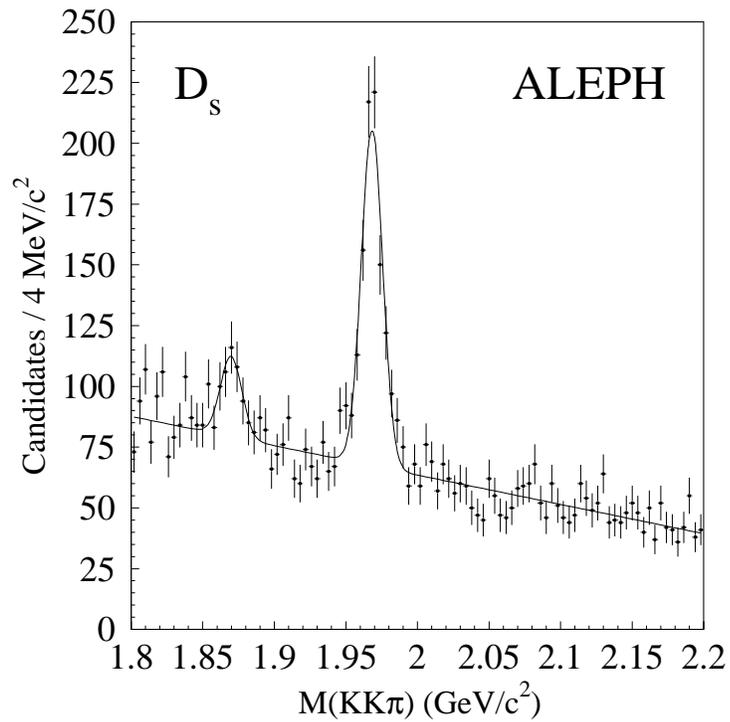,height=10.5cm}}
    \caption{{\small $\K^+\K^-\pi^+$ invariant mass distribution, with 
the fit result. The peak at low mass comes from the $\dpl$}}
\label{ds}
\end{figure}

\unboldmath
The $\dsp$ is reconstructed in the decay chain 
$\dsp \rightarrow \phi \pi^+$ with a
subsequent decay $\phi \rightarrow \K^+ \K^- $. The kaon momenta must
exceed $1.5~\gevc$ and the pion momentum must be greater than $2.5~\gevc$.
The kaon candidates are required to have $\dedx$ information available
and to fulfill $\chi_{\K} + \chi_{\pi} < 1$.
In addition all tracks for which $\dedx$ is available must satisfy 
$|\chi_x| < 2.5$ for the given particle hypothesis.
If $|M_{\K \K}-M_{\phi}| < 5~\gevcc$ and $p_{\phi} > 4~\gevc$ the candidate is
kept and the three particles fitted to a common vertex.
Candidates with  a $\chi^2$ greater than 15 (there are 3
degrees of freedom)
are rejected.
A cut on $|\cos\lambda^*| > 0.4$, where $\lambda^*$ is the angle
between the kaon and the pion in the $\phi$ rest frame, motivated by the
$P$-wave nature of the $\phi$ decay, strongly suppresses the background.
The mass distribution of the $\dsp$ candidates is shown in  
Fig.~\ref{ds}.

\boldmath
\subsection{$\dsstp$ Meson}
\unboldmath

The $\dsstp$ meson is reconstructed in the decay channel $\dsstp
\rightarrow\dsp\gamma$, followed by $\dsp\rightarrow \K^+\K^-\pi^+$ through
either the $\phi\pi^+$ or the ${\bar{\K}}^{*0}\K^+$ decay mode.

In the $\dsp$ reconstruction, tracks are taken as kaon candidates when
the $\dedx$ measurement satisfies $|\chi_{\K}|<|\chi_\pi|$, and as pion
candidates when $|\chi_{\K}|>|\chi_\pi|$. To select the $\dsp$
mesons decaying into $\phi\pi^+$, pairs of opposite-charge kaon candidates 
satisfying $|M_{\K \K}-M_\phi|<8\ \mevcc$ are
required to form a vertex with a pion candidate with a probability
greater than $1\%$ and $|\cos\lambda^*|$ greater than 0.4.
$\dsp$ candidates decaying into 
${\bar{\K}}^*\K$ are selected when kaon and pion
candidates with opposite charges fulfill $|M_{\K\pi}-M_{\K^*}|<25\ \mevcc$
and $|\cos\lambda^*|>0.6$.

\begin{figure}[htbp]\centering
  \setlength{\unitlength}{1.0mm}
  \begin{picture}(150,150)
    \put(-12,-50){
    \mbox{\epsfig{file=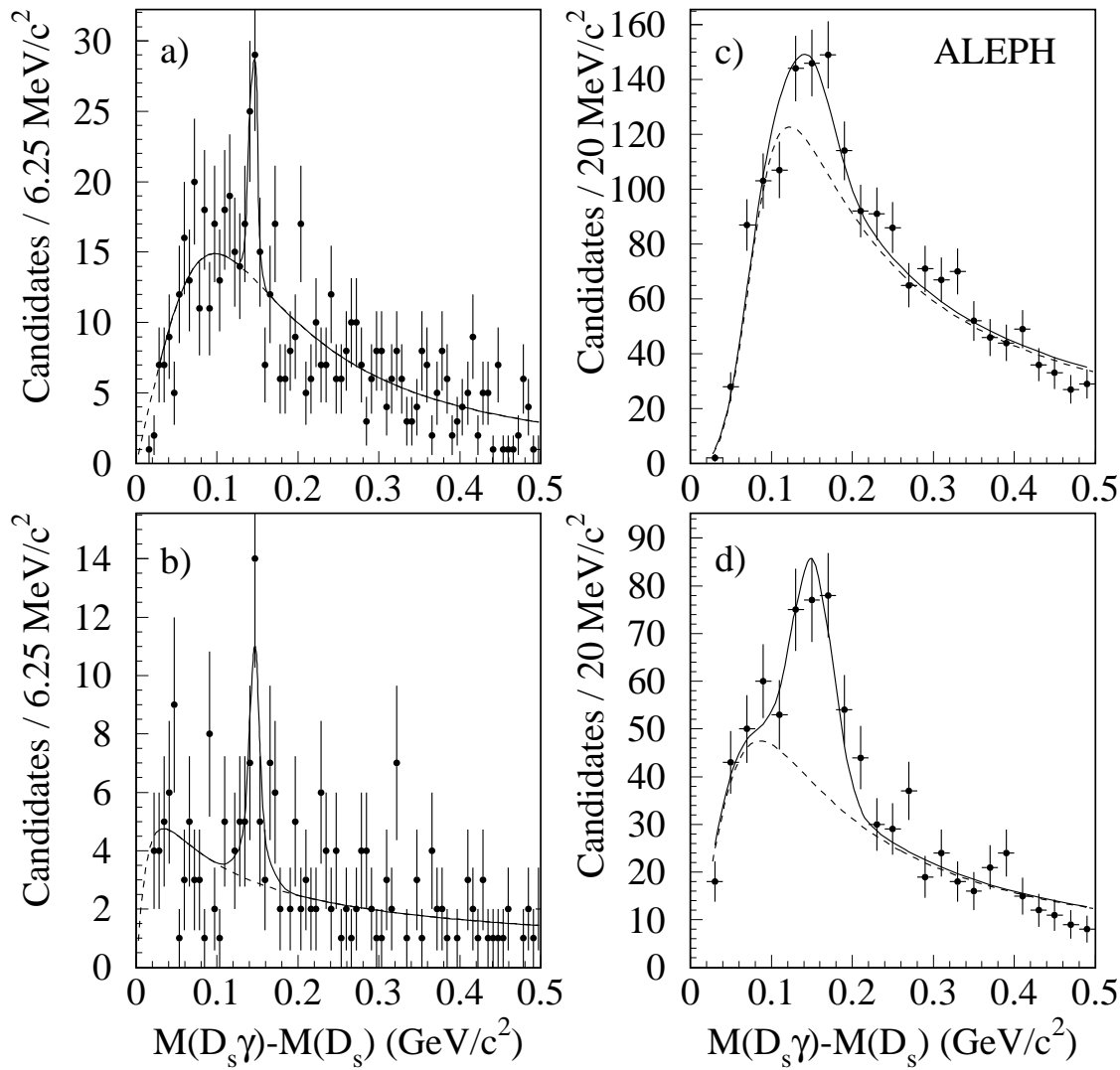,height=17cm}}}
  \end{picture}
    \caption{{\small $M(\dsstp)-M(\dsp)$ distributions with conversions in the
      $\bb$-enriched sample (a) and in the $\cc$-enriched sample (b), and
      with calorimetric photons (c), (d). The binnings have been chosen
      to match the different resolutions in the two photon detection methods.
      The dashed line is the background fit described in Section~6}}
\label{dsstar}
\end{figure}

Photon candidates for the $\dsstp$ are selected
when they have an energy greater than 0.6 GeV 
in the electromagnetic calorimeter and do not form a $\pi^0$ candidate.
Photon conversions in the detector are identified by searching
for electron tracks that are geometrically consistent with the conversion
hypothesis. To increase the efficiency for converted photons,
electron-positron pairs where one track is not detected are also identified,
as described in \cite{bstar}. Photon conversion
candidates must have an energy larger than 0.3
GeV to be taken into account.

To be able to measure separately the production rate of $\dsstp$ mesons
in $\cc$ and $\bb$ events, two different subsamples of $\dsstp$ candidates
are selected. One subsample is enriched in $\bb$ events and consists of 
$\dsstp$ candidates with a $\dsp$ momentum 
$8\ \gevc<p_{\D_{\rm{s}}}<20\ \gevc$,
a kaon momentum $p_{\K}>1.5\ \gevc$, a pion momentum $p_\pi>1\ \gevc$,
a fitted decay length $\delta \ell>500\ \mu$m, and a photon energy
$E_\gamma<3$ GeV. The $\bb$ purity of this $\dsstp$ signal is $85\%$.
In the other
subsample, enriched in $\cc$ events, the requirements are 
$p_{\D_{\mathrm{s}}}>20\ \gevc$,
$p_{\K}>2.5\ \gevc$, $p_\pi>1.5\ \gevc$, $\delta \ell>0$, and $E_\gamma<5$ GeV,
corresponding to a purity of the $\dsstp$ signal of about $60\%$ in 
$\cc$ events.
In Fig.~\ref{dsstar} the $M(\D_{\mathrm s}^*)-M(\D_{\mathrm s})$ 
distributions with 
calorimetric photons and conversions are plotted.

\boldmath
\subsection{$\lc$ Baryon} 
\unboldmath
The $\lc$ is observed via the decay mode $\lcpkpi$. To form a $\lc$ 
candidate three
tracks are required, each having at least one vertex detector hit.
The momenta must be 
greater than $5.0~\gevc$, $3.0~\gevc$ and $1.0~\gevc$ for the proton, 
kaon and pion candidates respectively and the total energy must exceed 
$0.2 \times \eb$. 
 For particle identification, 
$|\chi_h| < 2.5$ is required for the given particle hypothesis and 
two pion veto cuts are applied: $\chi_\pi < -2.0$ for the proton
candidate and $\chi_\pi < -1.0$ for the kaon candidate.
 
\begin{figure}[htbp]\centering
    \mbox{\epsfig{file=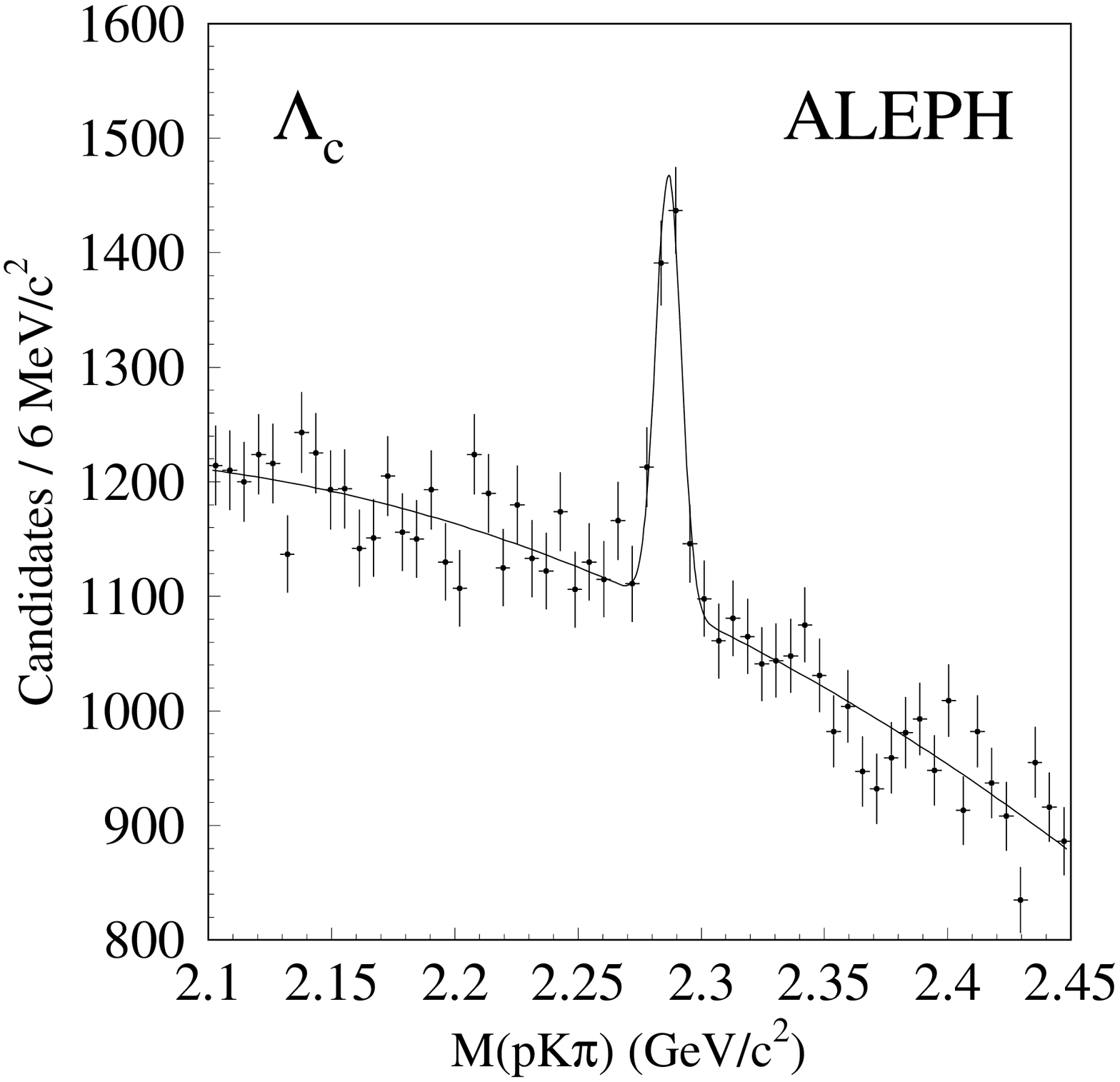,height=9cm}}
    \caption{{\small $\p\K^-\pi^+$ invariant mass distribution with the fit 
result.}}
\label{lamc}
\end{figure}

After a $\chi^2$ fit of the tracks to a common vertex the $\chi^2$ 
probability
has to be greater than $1\%$. 
The mass distribution of the $\lc$ candidates is shown in Fig.~\ref{lamc}.

\boldmath
\section{$\dstp$ Spectrum and Charm Fragmentation}
\begin{figure}[htbp]\centering
  \setlength{\unitlength}{1.0mm}
  \begin{picture}(120,120)
 \put(-5,-35){
    \mbox{\epsfig{file=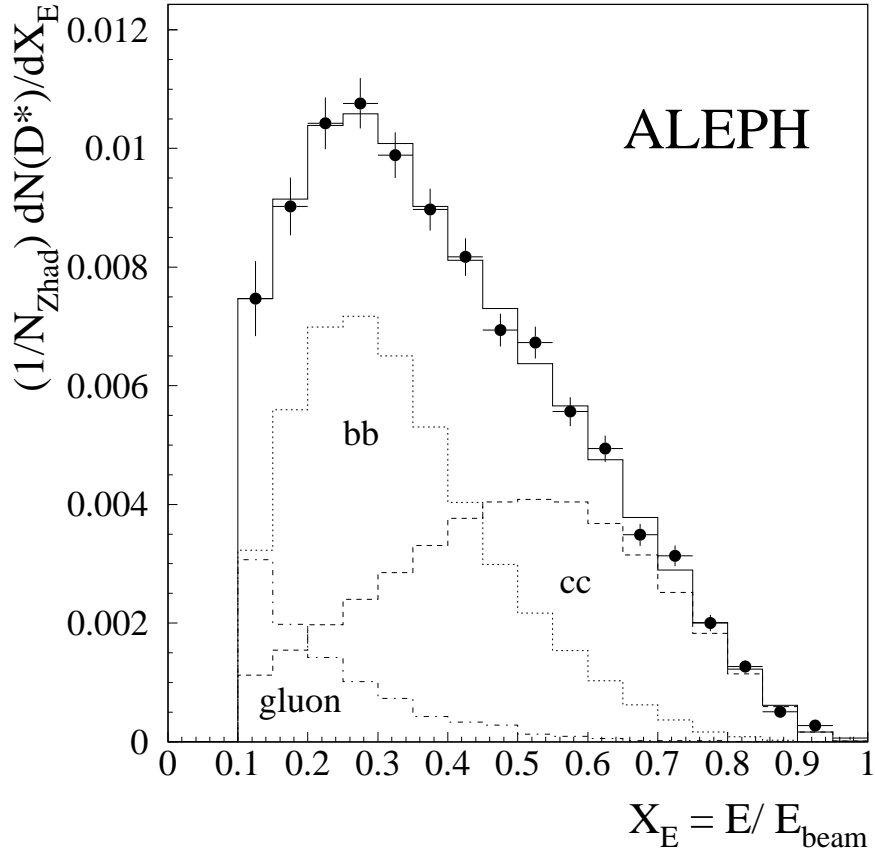,height=13cm}}
}
\end{picture}
    \caption{{\small Distribution of $\xe$ for the $\dstpm$'s. 
The data (points with error bars) are compared to the fit described
in the text. The three contributions: $\bb$ (dotted line), 
$\cc$ (dashed line) and
gluon splitting to heavy quarks (dashed-dotted) are shown.}}
\label{xe}
\end{figure}

\unboldmath
Because of its large signal/background ratio,
the $\dstp$ sample, selected as described in section 3.2,
is used to measure the fractional energy spectrum.
The $\xe$ range between 0.1 and 1.0 is divided into 18 equal bins.
In each bin, an analytical form is fitted outside the $\dstp$ peak
to the mass-difference distribution, so that
the background can be subtracted bin by bin.
The efficiency-corrected $\xe$ distribution is
shown in Fig.~\ref{xe} and the numerical values of the production 
rates are given in Table~\ref{td0}.

\begin{table}[htbp]
\caption{{\small Differential $\dstpm$ production rate as a function 
of the fractional energy $\xe$. The errors are statistical only.}}
\begin{center}
\begin{tabular}{|c|c|}
\hline
$\xe$ range & 1/N(Z$\to$hadrons)$\times$(dN($\dstpm$)/d$\xe$) \\
 & ($10^{-3}$) \\

\hline
0.10-0.15 & $7.47\pm0.63$ \\
0.15-0.20 & $9.03\pm0.49$ \\
0.20-0.25 & $10.42\pm0.44$ \\
0.25-0.30 & $10.76\pm0.43$ \\
0.30-0.35 & $9.89\pm0.38$ \\
0.35-0.40 & $8.97\pm0.35$ \\
0.40-0.45 & $8.17\pm0.32$ \\
0.45-0.50 & $6.94\pm0.28$ \\
0.50-0.55 & $6.73\pm0.27$ \\
0.55-0.60 & $5.56\pm0.24$ \\
0.60-0.65 & $4.94\pm0.22$ \\
0.65-0.70 & $3.49\pm0.18$ \\
0.70-0.75 & $3.13\pm0.17$ \\
0.75-0.80 & $2.00\pm0.14$ \\
0.80-0.85 & $1.27\pm0.11$ \\
0.85-0.90 & $0.50\pm0.07$ \\
0.90-0.95 & $0.27\pm0.05$ \\
0.95-1.00 & $0.06\pm0.03$ \\
\hline
\end{tabular}
\end{center}
\label{td0}
\end{table}

\begin{figure}[htbp]\centering
  \begin{center}
  \setlength{\unitlength}{1.0mm}
  \begin{picture}(120,120)
 \put(0,0){
    \mbox{\epsfig{file=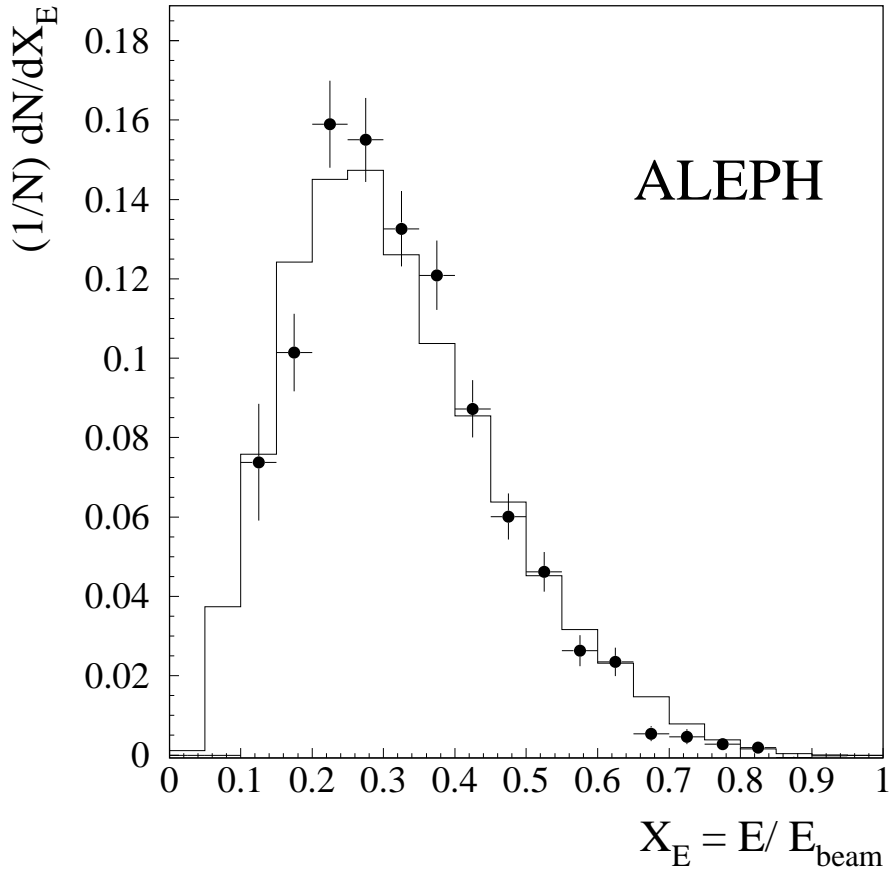,height=13cm}}
}
\end{picture}
    \caption{{\small Distribution of $\xe$ for the $\dstp$'s from the $\bb$
source compared with the Monte Carlo prediction.
The data (points with error bars) are obtained by applying the 
lifetime-mass btag on the hemisphere opposite to the $\dstpm$, and
corrected for efficiency. The remaining charm contribution is 
subtracted.}}
\label{bshape}
  \end{center}
\end{figure}

The $\dstp$ spectrum is
interpreted as the sum of three contributions: hadronisation
of a charm quark produced in a $\Zcc$ decay, decay of a bottom hadron
from a $\Zbb$ decay, and gluon splitting into a pair of heavy quarks
which hadronise or decay into a $\dstp$. 
The $\xe$ distribution of the three contributions are taken 
from the Monte Carlo simulation which uses
the Peterson {\it et al.} fragmentation scheme~\cite{peterson} 
for heavy quarks.
The $\bb$ contribution is compared with data using a high purity b-tag      
in the opposite hemisphere~(Fig.~\ref{bshape}).
A possible contamination from two-photon production
has been investigated by a Monte Carlo study normalised to
data \cite{gamgam} and found to be negligible ($<0.25 \%$).

To describe the production mechanism, three parameters are fitted to
the data : the fraction $f_g$ of $\dstp$ from gluon splitting, the 
fraction $f_b$ of $\bb$ events in the remaining (no gluon splitting)
sample, and the Peterson fragmentation parameter 
$\varepsilon_c$.  The fit is carried out in two steps iteratively.
Firstly a two-dimensional fit to the distribution
($\xe$,${\cal B}_{\rm{tag}}$), where ${\cal B}_{\rm{tag}}$ is the tagging
variable described in Section~2 is performed in order to
constrain $f_b$.
$\xe$ is required to be larger than 0.25 
to reduce to a negligible level the gluon splitting contribution.
The shape of the c and b contributions are taken from a full
Monte Carlo simulation. The small gluon splitting contribution is
taken in size and shape from the Monte Carlo simulation. The 
combinatorial background distribution is taken from the side band
of the $\K\pi\pi - \K \pi$ mass difference in the data sample.
The ($\xe$,${\cal B}_{\rm{tag}}$) plane is 
segmented into 62 bins of variable size in such a way that they all
contain enough statistics for a $\chi^2$ fit of $f_{\b}$
to be performed.

In a second step, fixing
$f_{\b}$ to this value, $f_{\g}$ and
$\varepsilon_{\ch}$ are extracted from
a fit to the $\xe$ distribution using the function
\begin{equation}
\label{xefiteq}
(1-f_{\g}) 
\left[ f_{\b} \frac{1}{N_{\bb}}\frac{dN^{\bb}}{d\xe}+(1-f_{\b})
\frac{1}{N_{\cc}}\frac{dN^{\cc}}{d\xe} \right]  
+ f_{\g} \frac{1}{N_{\glucc}}\frac{dN^{\glucc}}{d\xe}.
\end{equation}

\begin{figure}[htbp]\centering
  \begin{center}
  \setlength{\unitlength}{1.0mm}
  \begin{picture}(120,120)
 \put(0,0){
    \mbox{\epsfig{file=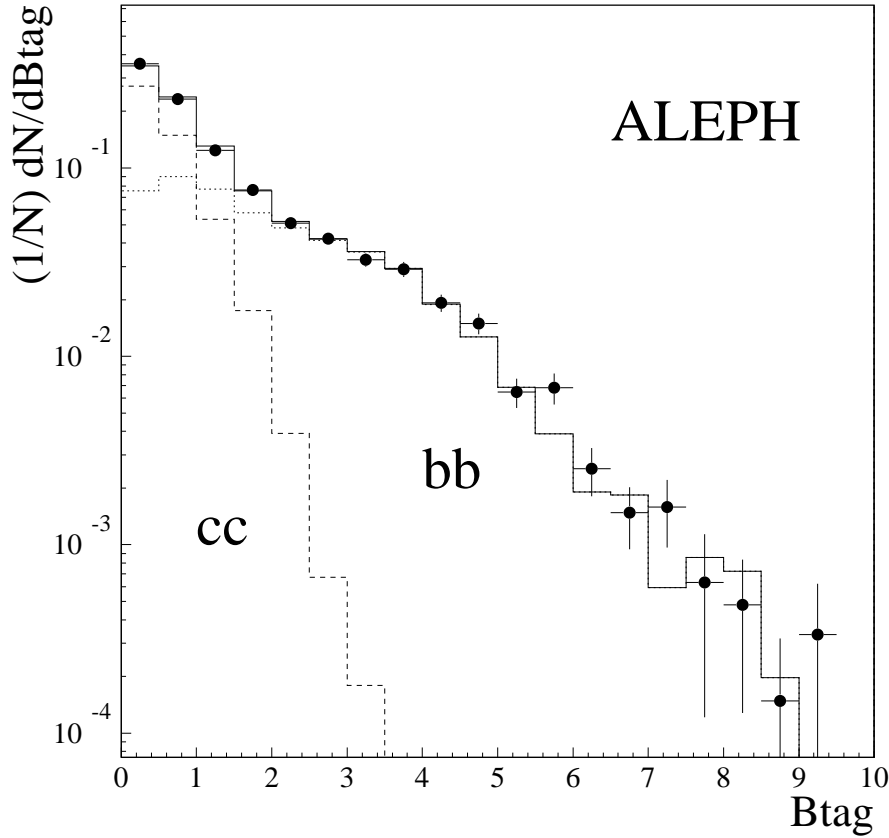,height=13cm}}
}
\end{picture}
    \caption{{\small Distribution of the b-tagging variable for the $\dstpm$'s
with $\xe > 0.25$ 
The data (points with error bars) are compared with the 
projection onto the ${\cal B}_{tag}$ axis of the fit described
in the text. The two contributions: $\bb$ (dotted line) 
and $\cc$ (dashed line) are shown.}}
\label{btag}
  \end{center}
\end{figure}

%
The two steps are iterated and once convergence is reached, a
minimal $\chi^2$ of the 2D fit of 71.3 for 61 degrees of freedom is obtained 
for $f_b=0.534\pm0.016$, corresponding
to $R_b \ f(b \to \dstpm)/R_c \ f(c \to \dstp) = 1.15\pm0.06$.
The uncertainty of this measurement is statistically dominated.
The fitted distribution of the ${\cal B}_{\rm{tag}}$ variable 
is shown in Fig.~\ref{btag}, compared with data.

The result of the 1D-fit is shown in Fig.~\ref{xe}; it has a $\chi^2$ of
11.9 for 14 degrees of freedom (the last three bins are grouped
together in the fit). 
The fitted value of $f_{\g}$ is related 
to the average number of gluon splitting events to charm per Z hadronic
decay and gives $\bar{n}_{\glucc} = (4.7 \pm 1.0 {\mathrm (stat.)})\%.$ 
The systematic error on this result however is large, due to the efficiency 
uncertainties at low $\xe$.
In the fit, the
JETSET~\cite{jetset} Monte Carlo shape for the gluon splitting and the Peterson
parameterisation for the heavy quark fragmentation 
functions~\cite{peterson} are assumed. In the next section, 
a more sensitive analysis dedicated to
this measurement is described.

The fitted value $\varepsilon_{\ch} = (33.9 \pm 3.7) \times 10^{-3}$
can be converted into a measurement of the average fractional
energy of the $\dstp$ in $\Zcc$ events of 
$${\langle \xe (\dst) \rangle }_{\cc} = 0.4878 \pm 0.0046 \mathrm{(stat.)}
\pm 0.0061 \mathrm{(syst.)} .$$

The quoted systematic uncertainty is dominated by the choice of
parameterisation of the fragmentation function, as in
Ref.~\cite{dspap}.

Using the Monte Carlo shape for the very small extrapolation to
the entire $\xe$ range (from 0.1 to the threshold at 0.044), 
the number of $\dstpm$ per hadronic Z decay in the mode considered 
is found to be
$$\bar{n}_{\dstpm} \times  B(\dstp \to \dz \pi^+) \times
  B(\dz \to \K^- \pi^+)
= (5.114 \pm0.067({\rm stat.}) \pm 0.072 ({\rm syst.}))
 \times 10^{-3}.$$ 
A $0.5\%$ uncertainty from the finite Monte Carlo statistics is 
included in the statistical uncertainty. This result includes the 
gluon splitting contribution. 
Subtracting this contribution and using the ratio of the
$\bb$ to the $\cc$ contribution quoted above,  
the probability for a c-quark to hadronise into a $\dstp$ is measured 
to be $$f(\ch \to \dstp) = 0.2333 \pm 0.0102 {\mathrm (stat.)}
\pm 0.0084 {\mathrm (syst.)}
\pm 0.0074 {\mathrm (B.R.).}$$
The normalisation of this result relies on the branching fraction
given in ref.~\cite{pdg}.

Systematic errors on these quantities arise from various sources.
The uncertainty from $f_{\b}$ has been evaluated by varying it 
by $\pm1$ standard deviation. 
The uncertainty on the efficiency has been estimated on a cut by cut
basis. The uncertainty from the mass and mass-difference cuts have
been assessed by varying the fitted mass and mass difference within
errors yielding a $0.5\%$ and $0.3\%$ contribution to the relative
error on the efficiency.
The Prob($\chi^2$) cut efficiency has been compared between data
at high $\xe$ and Monte Carlo, and found to differ by $8.3 \% $. 
This is taken as the uncertainty on the efficiency of this cut
and results in a $0.6\%$ uncertainty
on the overall normalisation. 
To assess the uncertainty from the cut on decay length significance,
the resolution on this quantity has been varied by $10\%$. The
resulting relative change in efficiency is found to be $(1.6\pm0.6)\%$
for the $\delta \ell / \sigma > 0 $ cut and $(0.06\pm0.90)\%$ for
the $\delta \ell / \sigma > 1$ cut. Taking a $1.6\%$ uncertainty
in the range $0.3 < \xe < 0.4$ and $0.9\%$ for $\xe$ below 0.3,
the resulting uncertainty on the overall normalisation is $0.4 \%$.
The probability of nuclear interaction in
the detector is simulated at the $10 \%$ level,
determined from a study of tau pairs where the pion from one
of the taus undergoes such an interaction in the detector. 
This yields a $0.7\%$
relative error on the efficiency.
For the systematics from background subtraction, the analytical
shape of the background has been modified and the fit range for
the mass-difference fit has been varied.
This results in a $0.2\%$ overall error on the absolute 
normalisation, localised in the first three $\xe$ bins where the 
background is large (the uncertainty reaches $4.5\%$
in the first bin). Summing linearly the errors from the P($\chi^2$) cut
and the $\delta \ell / \sigma$ cut, 
as they both 
pertain to vertexing and are strongly correlated, and then summing
in quadrature all the contributions, 
the overall normalisation uncertainty is
$1.4\%$. Details are given in Table~\ref{td1}.
As a check, the whole analysis has been repeated with the
decay length significance cut
relaxed by 1 unit and the cut in $\chi^2$ probability for the $\dz$
vertex removed. The final number of events changes by $1\%$.

\begin{figure}[htbp]\centering
  \setlength{\unitlength}{1.0mm}
  \begin{picture}(120,120)
 \put(0,0){
    \mbox{\epsfig{file=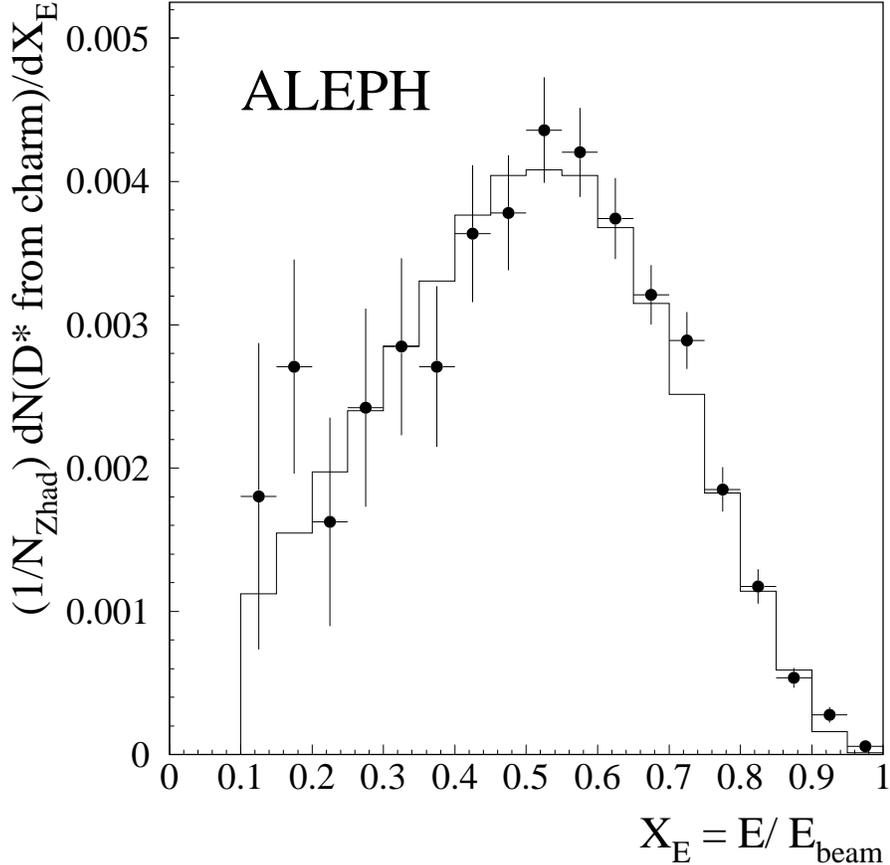,height=13cm}}
}
\end{picture}
    \caption{{\small 
Distribution of $\xe$ for the $\dstpm$'s from the $\cc$ source. 
The data (points with error bars) are compared to the Monte Carlo
simulation.}}
\label{xec}
\end{figure}

Some of these systematic effects are $\xe$ dependent. They have been
propagated to the gluon splitting measurement and the average $\xe$
measurement.

As a check on the stability of the result, the fit has been repeated with   
$f_{\b}$ left free. The fitted parameters are perfectly consistent
with the result of the whole procedure, but the uncertainty on $f_b$ is 
increased by a factor of 1.7. 
 
In Fig.~\ref{xec}, the spectrum of the $\dst$s from charm is shown. This 
spectrum (points with error bars) is obtained from the data points
of Fig.~\ref{xe} or Table~\ref{td0}, from which the b contribution
and the gluon splitting contribution are subtracted. The b contribution
is extracted from the b-tagged data  
of Fig.~\ref{bshape}, normalised using the fitted $f_{\b}$. The Monte
Carlo shape, normalised to the measurement presented in the next section,
is used for the gluon-splitting contribution.

\begin{table}[htbp]
\caption{{\small 
Relative systematic errors in percent on the $\dstpm$ production rate 
measurement.
Some of them are strongly $\xe$ dependent (see text). 
The two vertexing uncertainties are considered as fully correlated 
and added linearly.}}
\begin{center}
\begin{tabular}{|c|c|}
\hline
Source & $\delta \epsilon / \epsilon $ \\
\hline
$\Delta M $ cut & $0.3\%$ \\
$M(\K \pi)$ cut & $0.5\%$ \\
P($\chi^2$) cut & $0.6\%$ \\
$\delta \ell / \sigma$ cut & $0.4\%$ \\
nuclear interactions & $0.7\%$ \\
\hline
Backg. subtraction & $\delta N/N = 0.2 \%$.\\
\hline
Total systematics & $ 1.4\%$ \\
\hline
\end{tabular}
\end{center}
\label{td1}
\end{table}

\boldmath
\section{Measurement of the Gluon Splitting Rate to $\cc$}
\begin{figure}[htbp]\centering
 \setlength{\unitlength}{1.0mm}
  \begin{picture}(90,90)
 \put(0,0){
\mbox{\epsfig{file=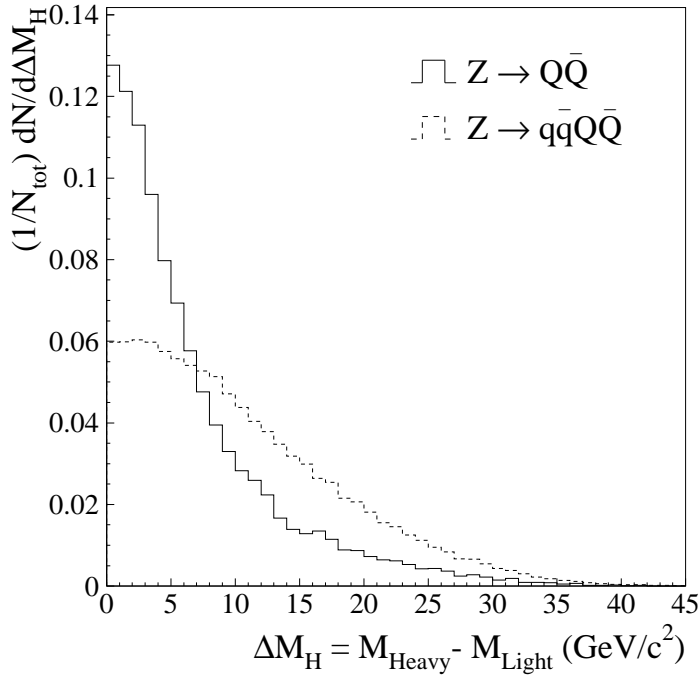,height=10cm}}
}
\end{picture}
\caption[Hemisphere mass difference]
        {\small
        {Difference of heavy and light hemisphere masses in Z decays
        with (dashed line) and without (full line) gluon splitting to 
        heavy quarks (from JETSET). The events entering these 
        distributions are required to contain at least one $\dstpm$.
        Q stands for b and c, and q for all flavours. 
        \label{fig:hemi_mass}}}
\end{figure}

\unboldmath
In order to determine more precisely the rate \ngcc one
more variable, related to the event topology, has been used.
In a \gcc event the two quarks from the gluon splitting, together with one of
the primary quarks, tend to end up in the same hemisphere, 
resulting in a large invariant mass for this hemisphere and a 
relatively light opposite hemisphere.
Denoting by  $M_{\rm{heavy}}$ ($M_{\rm{light}}$) the invariant mass
of the heavier (lighter) hemisphere in an event, the difference
$\Delta M_H = M_{\rm{heavy}}-M_{\rm{light}}$ has been used to discriminate
between D$^*$'s produced from primary quarks and from gluons.
The distributions of $\Delta M_H$, as given by the JETSET 
event generator
\cite{jetset}, are shown in Fig.~\ref{fig:hemi_mass}.
In JETSET $\dst$'s from primary quarks are nearly
evenly shared among the hemispheres while $\approx 75\%$ of the $\dst$'s 
from gluons are found in the heavy hemisphere.

The sample of reconstructed $\dstp$ mesons is divided into two samples
depending upon whether the $\dstp$ is in the
light or heavy hemisphere. A two--dimensional binned
likelihood fit is then performed to the distributions in the $X_E/\Delta M_H$
plane of these two samples using the function :

\begin{equation}
(1-f_{\g}) 
\left[ f_{\b} \frac{1}{N_{\bb}}\frac{d^2N^{\bb}}{d\xe d\Delta M_H}+
(1-f_{\b})
\frac{1}{N_{\cc}}\frac{d^2N^{\cc}}{d\xe d\Delta M_H} \right] 
\hfill \cr \hfill
+ f_{\g} \frac{1}{N_{\glucc}}\frac{d^2N^{\glucc}}{d\xe d\Delta M_H}.
\end{equation}

For this analysis the event selection has been tightened in order to ensure
that the event is fully contained in the detector: The total energy
$E_{\rm{ch}}$
of all charged tracks is required to exceed $15\,\mbox{GeV}$ and events
strongly unbalanced in momentum projected along the beam axis $p_z$
are rejected by
requiring $|\sum p_z| / E_{\rm{ch}} < 40\%$. This selection has an efficiency
of $89.3\%$ for hadronic Z decays without the process \gQQ and $91.9\%$ for
\gQQ events (Q stands for b or c), as estimated from Monte Carlo.

\begin{figure}[htbp]\centering
\mbox{\epsfig{file=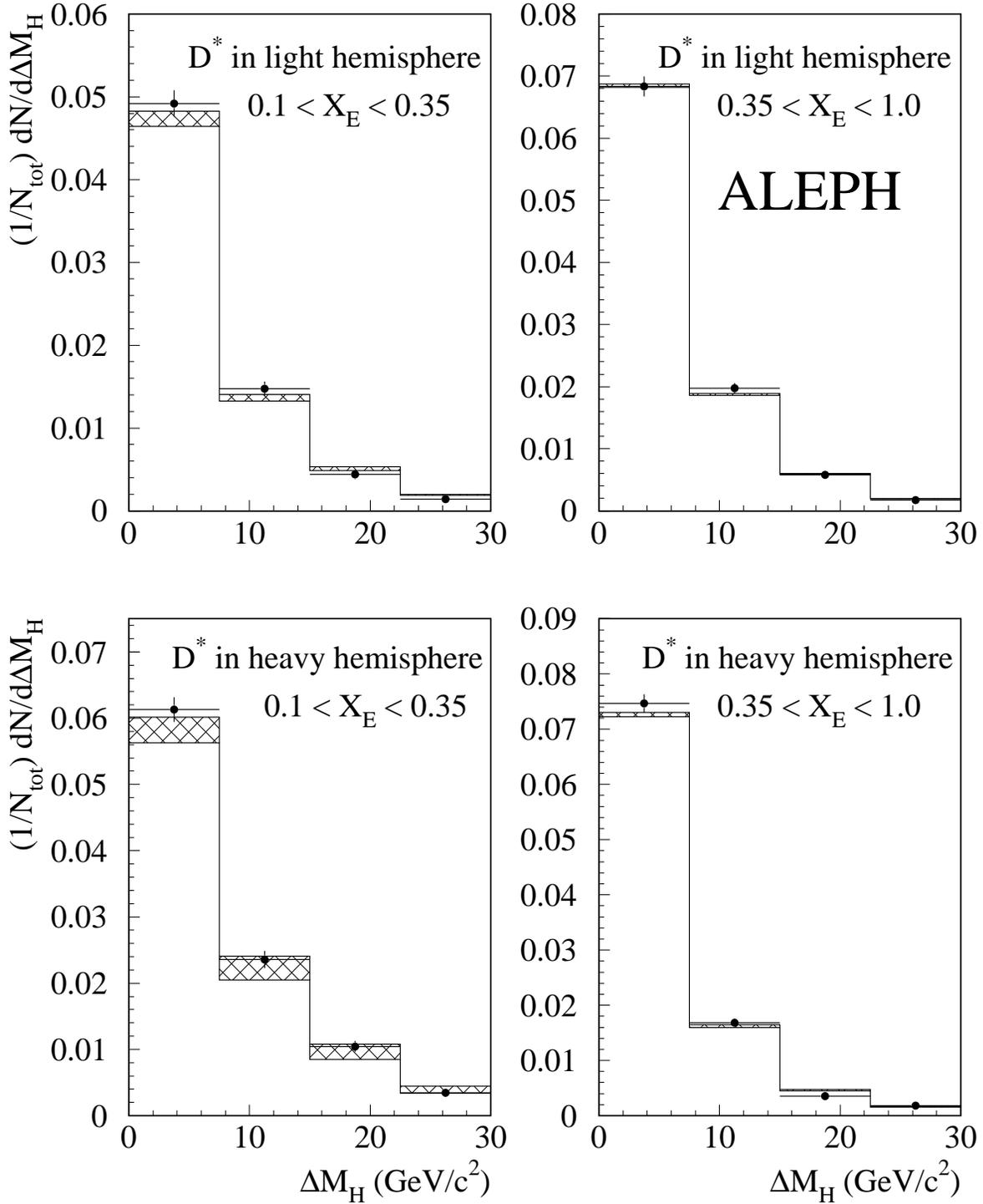,height=21.5cm}}
\caption[Fit in four bins]
        {{\small
        Hemisphere mass difference distributions.
        The data (points) are subdivided into four bins with D$^*$ in
        the heavy/light hemisphere and for low/high $X_E$. The fitted
        contribution from primary quarks is shown as the solid histogram,
        the \gQQ contribution is added as cross-hatched area.}
        \label{fig:xe_dm_fit}}
\end{figure}

\begin{figure}[htbp]\centering
 \setlength{\unitlength}{1.0mm}
  \begin{picture}(90,90)
 \put(0,-5){
\mbox{\epsfig{file=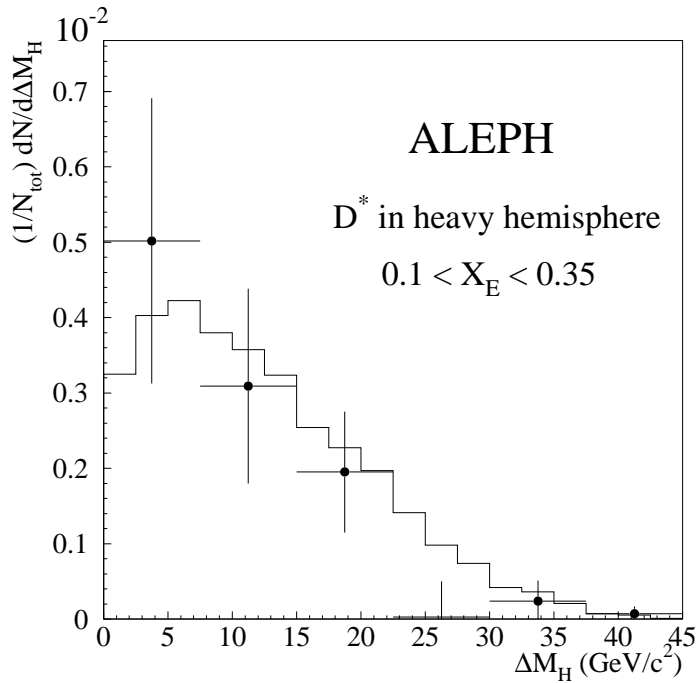,height=10cm}}
}\end{picture}
\caption[Fit of b/c subtracted distribution]
        {{\small Comparison of the data with the fitted contribution from 
          primary quarks subtracted (points) and the predicted gluon 
          splitting distribution (histogram).}
        \label{fig:dm_bcsub}}
\end{figure}

 The $\dstp$ mesons are reconstructed with cuts similar to the ones used in
section~3. The hemispheres are defined by a plane perpendicular to the
thrust axis and their invariant masses are calculated using 
energy flow objects with an energy above $1\,\mbox{GeV}$.
Again, the analysis proceeds in two steps.
A fit to the one--dimensional ${\cal B}_{\rm{tag}}$ distribution yields the 
fraction $f_{{\cal B}tag} = 0.524 \pm 0.015$ of D$^*$ from b hadron decays
in the 
selected sample. The error includes the systematic uncertainties but is 
dominated by the statistical error. 
Then the two two-dimensional distributions $X_E/\Delta M_H$, 
depending upon whether the D$^*$ is in the 
heavy or light hemisphere, are fitted simultaneously 
using a binned likelihood fit ($18 \times 18$ bins).
The b-contribution excluding gluon splitting $f_{\b}(1-f_{\g})$
is constrained to $f_{{\cal B}tag}$ by a $\chi^2$ term.
The result of the fit is 
\begin{displaymath}
  f_{\g} = 0.0522 \pm 0.0077\,,\,f_{\b} = 0.5308 \pm 0.0098
\end{displaymath}
from which \ngcc is computed as
\begin{eqnarray}
  \label{eqn:ngcc}
   \ngccm & = & R_{\ch} \frac{\epsilon_{\ch}}{\egcc}\frac{1}{1-f_{\b}}
         \left[\frac{f_{\g}}{1-f_{\g}} - \frac{1}{R_{\b}} 
         \frac{\egbb}{\epsilon_{\b}}
          f_{\b} \ngbbm \right] \\
        & = & (3.23 \pm 0.48(\mbox{stat.}))\%\;.
\end{eqnarray}
Here, $\epsilon_{i}$ is the efficiency to reconstruct a $\dstp$ from a given
source $i$. 
The term in brackets provides a $\sim 10\%$ correction for the 
\gbb contribution which is taken from ref.~\cite{gsplitb}:
$\ngbbm = (2.77 \pm 0.42 \pm 0.57 ) \times  10^{-3}$.
In Fig.~\ref{fig:xe_dm_fit} projections of the two-dimensional 
distributions
on $\Delta M_H$ are shown for the two hemispheres for two ranges in $X_E$.
The agreement between data and Monte Carlo is good, in particular it can be
seen that the high-$\xe$ data are well described without the gluon splitting
contribution but that its inclusion is necessary for $X_E<0.35$ in the
heavy hemisphere. The $\chi^2$ probability for this distribution degrades 
from 0.40 to 0.001 for the four bins  
when the gluon splitting contribution is neglected. In order to show more
clearly the shape of the gluon splitting signal in this region, 
Fig~.\ref{fig:dm_bcsub} shows the data with the fitted 
contribution from primary quarks
subtracted. Good agreement with the Monte Carlo prediction is observed.

Systematic errors have been studied for the efficiency of D$^*$ reconstruction,
the fitting procedure, the modelling of the Monte Carlo distributions, and
the parameters entering Eq.~(\ref{eqn:ngcc}). They are summarised in 
Table~\ref{tab:ngcc_syst}.\\
\begin{table}[h]
\caption [Systematic errors]{\label{tab:ngcc_syst}
         {{\small List of absolute systematic errors on \ngcc}}}
\begin{center}
\begin{tabular}{|l|r|} \hline
        Source & $\Delta$\ngcc $(10^{-2})$\\ \hline
        $\dstpm$ reconstruction   & $ 0.25$ \\ 
        Monte Carlo statistics & $ 0.19$ \\ 
        b--contribution        & $ 0.09$ \\ 
        Hemisphere selection   & $ 0.15$ \\ 
        $\varepsilon_c$        & $ 0.15$ \\ 
        $\varepsilon_b$        & $ 0.19$ \\ 
        $X_E(\mbox{b}\rightarrow \mbox{D}^*)$ from data
                               & $ 0.15$ \\ 
        D$^{**}\rightarrow\mbox{D}^*\,X$ & $0.04$ \\ 
        $\Lambda_{PS}$         & $ 0.15$ \\ 
        Hard gluon             & $ 0.08$ \\ 
        {\sc Herwig}           & $ 0.11$ \\ 
        {\sc Ariadne}          & $ 0.21$ \\ \hline
        Total                  & $ 0.53$ \\ \hline
\end{tabular}
\end{center}
\end{table}
\noindent
The errors on the number of reconstructed $\dstp$'s have been estimated as in
the previous section by comparing the efficiencies of the cuts between
Monte Carlo and data and varying them within their uncertainties. The
normalization of the background has also been changed within the errors.

The finite Monte Carlo statistics has been accounted for by repeating
the fit with the distributions randomly smeared. The fit has also been
redone not applying the $\chi^2$ constraint on the b contribution, and
leaving free the relative fractions of D$^*$'s in the two hemispheres 
for the c and b contributions separately.

Several parameters affecting the shape of the Monte Carlo distributions 
have been varied: $\varepsilon_\ch$ has been changed within the errors derived
from the above measurement and $\varepsilon_\b$ within
$\varepsilon_{\b} = 0.0045\pm0.0009$. The $X_E$ shape for
$\mbox{b}\rightarrow \mbox{D}^{*}$ as derived from the b--tagged sample has 
been used for the fit. The dependence of the fraction of D$^*$
mesons originating from a D$^{**}$ has been estimated by changing this rate
by $\pm 30\%$ for b and c separately.

The effects of gluon emission in the parton shower have been assessed by
changing the JETSET parameter $\Lambda_{PS}$ such that it corresponds to a
$\pm 4\%$ variation in $\alpha_s(M_{\rm{Z}})$. 
The rate of events in which both primary
quarks recoil against a hard gluon and populate the same hemisphere has
been varied by $30\%$. Finally, the $\Delta M_H$ shapes as
given by HERWIG 5.8~\cite{herwig} and ARIADNE 4.08~\cite{ariadne}
have been used.
For ARIADNE, 
the transverse momentum of a splitting gluon was required 
to be larger than its virtuality, as suggested in
Ref.~\cite{glubc}. ARIADNE yields the largest discrepancy and
this has been taken as the systematic error on the Monte Carlo model.
Adding all systematic errors quadratically, the average fraction of
hadronic Z decays where a gluon splits in a $\cc$ pair
is found to be
\begin{displaymath}
  \ngccm = (3.23 \pm 0.48(\mbox{stat.}) \pm 0.53(\mbox{syst.}))\%\;.
\end{displaymath}

\boldmath
\section{Measurement of the $\dsstpm$ production rate}
\unboldmath

The selection efficiencies for finding $\dsstpm$ mesons are evaluated from
the Monte Carlo using the mix of
$\ds\longrightarrow\phi\pi^\pm$ and $\dsstp\longrightarrow
{\bar{\K}}^{*0}\K^+$ decays predicted by the branching ratios in
\cite{pdg}. They are given in Table~\ref{tab:eff} for the selection
described in Section 3.5.
\begin{table}[htb]
    \caption{Selection efficiencies for $\dsstpm$ mesons (the errors
      are statistical only).}
  \begin{center}
    \begin{tabular}{|l|c|c|c|c|}\hline
      & \multicolumn{2}{c|}{conversions} &
      \multicolumn{2}{c|}{calorimetric photons} \\ \hline
      & \ & \ & \ & \\
      & $\Z\rightarrow \cc \ (\%)$ & $\Z\rightarrow \bb \ (\%)$ &
      $\Z\rightarrow \cc \ (\%)$ & $\Z\rightarrow \bb \ (\%)$ \\ \hline
      $\cc$-enriched & $0.38\pm0.04$ & $0.12\pm0.02$ &
      $2.72\pm0.12$ & $0.85\pm0.07$ \\ \hline
      $\bb$-enriched & $0.10\pm0.02$ & $0.29\pm0.03$ &
      $0.57\pm0.05$ & $1.60\pm0.09$ \\ \hline
    \end{tabular}
    \label{tab:eff}
  \end{center}
\end{table}

The number of events in the data samples was evaluated by fitting the
$\dsstp$ spectrum with the sum of two Gaussian functions for the signal and
a polynomial for the background. In the case of the calorimetric photons, 
the means and widths of the two Gaussians and the ratio of their area are 
taken from Monte Carlo. For converted photons the means of the two
Gaussians and the width of the broad one are left free in the fit.  
The background is parameterised to reproduce the Monte Carlo.
From the fit the number of events in the
$c\bar{c}$-enriched sample is $N_c=26\pm7$ ($139\pm22$) and in the
$b\bar{b}$-enriched is $N_b=36\pm11$ ($106\pm37$) with converted
(calorimetric) photons.
The production rates $f(c\rightarrow \dsstpm)$ and
$f(b\rightarrow \dsstpm)$ can be calculated by solving two equations
of the following form,
corresponding to the $\cc$ and $\bb$ enriched samples:
\begin{eqnarray}\nonumber
N_{\dsstp} & = & 2N_{\qq}\times B(\dsstp\rightarrow \dsp\gamma)
B(\dsp\rightarrow\phi\pi^+)B(\phi\rightarrow \K^+\K^-) \\
& & \ \, \times [\Rc f(\ch \rightarrow \dsstp) \epsilon_{\ch}+
\Rb f(\b \rightarrow \dsstpm) \epsilon_{\b}],
\end{eqnarray}
where $N_{\qq}$ is the number of hadronic events in the data sample,
$\epsilon_{\ch}$ ($\epsilon_{\b}$) the
selection efficiency for $\dsst$ mesons in $\cc$ 
($\bb$) events and $f(\ch \rightarrow \dsst)$
($f(\b \rightarrow \dsst)$) the probabilities of a c (b) quark
giving rise to a $\dsst$ meson. A similar formula can be written
for the $\D_{\rm{s}}^+\longrightarrow \K^{*0}\K^+$ channel. 
The results from the whole data sample are:
\begin{displaymath}
  \begin{array}{l}
    f(\ch \rightarrow \dsstp)=6.9\pm1.8\ {\rm (stat.)}\pm0.7\ {\rm (syst.)}
\pm1.8 {\rm (BR)}\% \\
    f(\b \rightarrow \dsstp)=11.3\pm3.5\ {\rm (stat.)}\pm1.0\ {\rm (syst.)}
\pm2.8 {\rm (BR)}\%\\
  \end{array}
\end{displaymath}
Here the last error is due to the final state branching ratios
taken from \cite{pdg}, while
the systematic error includes the uncertainty in the selection
efficiencies and background parameterisation in the fit. The latter
contribution is evaluated using different functional shapes for the
background; a further check is made by taking the background
shape evaluated bin by bin from Monte Carlo and using it in the fit.

\boldmath
\section{Production Rates of the Ground States }
\unboldmath
\boldmath
\subsection{Production of $\dz$ and $\dpl$}
\unboldmath
 
 The number of $\dz$ and $\dpl$ present in the samples are 
extracted from a fit to the invariant mass
distributions (Figs~\ref{d0},\ \ref{dp}), where the signal is parameterised by 
two gaussians with a common mean and the combinatorial
background by a polynomial function.
Resonant background contributions, such as $\kk$, $\dsp \to \phi \pi^+$,
$\dsp \to {\bar{\K}}^* \K$, and $\kp$ where the two mass assignments 
are reversed, are taken into account in the fit. 
The fractions of $\dz$ and $\dpl$ originating from the process $\Zcc$ are
measured by applying, on the hemisphere opposite to 
the reconstructed D meson,
the
lifetime-mass tag \cite{qmbtag} selecting b hemispheres with 99$\%$ purity. 
The charm fraction in the D meson samples is then given by   
$f_{\ch} = {{ \epsilon_{\Bb} - r_{\b-tag}  } 
\over {\epsilon_{\Bb} - \epsilon_{\Bc}}}$ 
where $r_{\b-tag}$ is the fraction 
of D mesons surviving the b-tag cut. 
The b hemisphere b-tag efficiency $\epsilon_{\Bb}$ is evaluated on data
using a double-tag technique
and $\epsilon_{\Bc}$, the c hemisphere b-tag efficiency, 
is obtained from Monte 
Carlo simulation~\cite{qmbtag}. Both efficiencies are corrected for the
selection bias induced by requiring the presence of a high momentum 
D meson in the opposite hemisphere. 
The fractions of $\dz$ and $\dpl$ mesons from charm
are found to be: 
 $f_{\ch}(\dz) =  0.787 \pm 0.019$ ; 
 $f_{\ch}(\dpl) = 0.797 \pm 0.020$, 
where the quoted errors are statistical.

 The small fraction of D mesons from 
  gluon splitting
 has been deduced using the measurement from  
section 5, the ALEPH measurement 
$\bar n_{\glubb}=(2.77\pm0.42\pm0.57) 
\times 10^{-3}$~\cite{gsplitb}, 
and the efficiency ratio for the selection of a D meson
${{\epsilon_{\g \rightarrow
\D}}\over{\epsilon_{\q \rightarrow \D}}} $ obtained from simulation.  
 
The number of fitted D mesons, together with the
charm fractions and the gluon splitting contamination, are reported in 
Table~\ref{t1}.
 
\begin{table}[htbp]
\caption{\label{t1}
{{\small $\dz$ and $\dpl$ selection: number of candidates; 
fraction of $\ch \to \D$ mesons in the sample;
efficiency of  $\ch \to \D$ meson reconstruction;
percentage of $\D$ mesons from gluon splitting.
All quoted errors are statistical.}}}
\begin{center}
\begin{tabular}{|c|c|c|c|c|}
\hline
  & & & & \\
  & N. candidates & Charm fraction & Efficiency & g$\to \QQ  (\%)$\\
  & & & & \\
\hline
$\dz$ &$7871\pm 129$ & $ 0.787 \pm 0.019$ &$0.200\pm 0.004$ & 0.8 \\
\hline
$\dpl$  & $7409\pm 153$ & $0.797 \pm 0.020$ &$0.191\pm0.002$ & 0.8 \\
\hline
\end{tabular}
\end{center}
\end{table}
  
From the numbers in Table~\ref{t1}, the following branching fractions are 
obtained:
$$ \Rc \times f(\ch \to \dz) \times B(\dz \to \K^- \pi^+) = 
 (0.370 \pm 0.011) \times 10^{-2} \\ $$
$$\Rc \times f(\ch \to \dpl) \times B(\dpl \to \K^- \pi^+ \pi^+) =
 (0.368 \pm 0.012) \times 10^{-2}  $$
 Different sources of systematic uncertainty have been investigated and are given in Table~\ref{t2}.
\begin{table}[htbp]
\caption{\label{t2}
{{\small 
Relative systematic errors in percent on the $\dz$, $\dpl$, $\dsp$ and
$\lc$ production rate measurements.}}}
\begin{center}
\begin{tabular}{|c|c|c|c|c|}
\hline
Source & $\dz$ ($\%$) & $\dpl$ ($\%$) & $\dsp$ ($\%$) & $\lc$ ($\%$) \\
\hline
 Nuclear interact. & 0.5 &  0.7 & 0.3 & 0.3\\
 $\dedx$ & 0.9 &  0.4 & 1.2 & 2.2 \\
 vertexing & 3.8 & 2.3 & 1.7 & 1.7 \\
\hline
 charm fraction & 0.7 &  0.7 & - & - \\
gluon splitting & 0.3 &  0.3 & 0.3 & 0.3 \\
fragmentation & 3.7 &  3.7 & 1.9 & 4.0 \\
charm physics & 1.6 & 2.1 & - & - \\
 fitting funct. & 1.5 &  1.5 & 4.0 & 1.6 \\
MC stat.& 2.0 &  1.0 & 3.0 & 2.0 \\
\hline
Total systematics & 6.2 &  5.3 & 5.8 & 5.5 \\
\hline
\end{tabular}
\end{center}
\end{table}
The first three errors are related to the detector performance. The fraction of tracks with $\dedx$ information
has been compared between data and Monte Carlo with a sample of candidates 
selected in the side-band regions of the $\dz$ and $\dpl$.
The Monte Carlo has been corrected for differences and the uncertainty 
on the ratio data/MC has been taken as a
systematic error. The $\dedx$ calibration and resolution 
have been studied on data for a pure kaon sample from the decay mode
$\dstkp$. Data/MC differences have been taken into account to obtain 
the $\dedx$ cut efficiency and the associated systematic error.
The fraction of D mesons forming a common vertex with a $\chi^2$-probability 
greater than $1\%$ is obtained from data,
from the fraction of selected candidates under the D meson mass peak, 
corrected by the fraction of background candidates 
 estimated on sideband regions. The resolution on the decay-length 
measurement has been checked on $\dz$ and $\dpl$ 
sideband samples, where most of the candidates carry no lifetime information. 
The difference found between data and Monte Carlo
has been taken as a systematic error for the decay length significance 
cut together with the error coming from the $\dz$ and $\dpl$
lifetime uncertainties.
 The fraction of tracks not reconstructed 
due to nuclear interaction in the $\kp$ and $\kpp$ decays 
is estimated from Monte Carlo to be 5~$\%$ and 7~$\%$ respectively, with
a systematic error of 10~$\%$.

The main systematic error on the determination of the charm fraction 
comes from the correction factor applied 
to the b-tag efficiencies. This arises from 
the selection of an 
energetic D meson in one hemisphere which
favours events with no gluon radiation, 
hence increasing the efficiency for the b-tag in the opposite hemisphere. 
 A comparison between data and Monte Carlo of the spectrum of the most energetic jet in the hemisphere opposite to the D meson 
has been performed to check the simulation of gluon emission. 
The quoted error on the correction factors arises  
from the limited Monte Carlo statistics and the 
difference between data and
Monte Carlo for gluon emission; this gives a relative 
error on the charm fraction 
of 0.7$\%$. The relative systematic error due to the
uncertainty on the  b-tag efficiencies is very small, of the order of 0.2$\%$. 
The systematics related to the charm fragmentation and the simulation of the decay chain leading to a $\dz$ and $\dpl$ are
evaluated by varying the fragmentation function and the fraction of charmed 
vector, pseudoscalar and excited meson (10 to $30\%$)
states produced in the charm hadronisation. The Peterson fragmentation 
function was varied to reproduce the measured 
$\dstp$ $\xe$ spectrum for different fractions of produced D meson states.
Different kinds of fit have been performed to assess the systematic error 
coming from the fitting.
For the signal function, the mean and width of the narrow gaussian 
distribution is left free, while the relative 
normalisation and width of the second gaussian is fixed to the 
Monte Carlo expectation. The fixed parameters have been
varied within 5$\%$ and the fit repeated with 
other background parameterisations (second-order polynomial, exponential). 
Fitting the background outside the 
mass peak and counting the number of candidates above the mass peak
 gives a consistent result.

The final result on the $\dz$ and $\dpl$ production rate measurements in 
$\Zcc$ is    
$$ \Rc \times f(\ch \to \dz) \times B(\dz \to \K^- \pi^+) = 
 (0.370 \pm 0.011 \pm 0.023) \times 10^{-2} \\ $$
$$\Rc \times f(\ch \to \dpl) \times B(\dpl \to \K^- \pi^+ \pi^+) =
 (0.368 \pm 0.012 \pm 0.020) \times 10^{-2}. $$
 
Dividing these numbers by the Standard 
Model value of $\Rc$ and the branching fraction for the
reconstruction mode, 
the following fractions are obtained:
$$f(\ch \to \dz) = 0.559 \pm 0.017 \pm 0.0035 \pm 0.013 {\mathrm(BR)}$$
and 
$$f(\ch \to \dpl) = 0.2379 \pm 0.0077 \pm 0.0129 \pm 0.0190 {\mathrm(BR)}.$$

 The consistency between the rate measurements  
is investigated.
Under the assumption that the observed difference between 
$f(\ch \to \dz)$ and $f(\ch \to \dpl)$ is only due to $\dst$ production,
i.e. if primary $\D$'s and $\dst$'s are produced evenly in the two charge 
states, the following
ratio must equal unity:
$${f(\ch \to \dz) -f(\ch \to \dpl)} \over  {2 \ f(\ch \to \dstp) \times
B(\dstp \to \dz \pi^+)} $$
This ratio is measured to be $1.02 \pm 0.12$.

\boldmath
\subsection{Measurement of the $P_V$ Ratio}
\unboldmath

The ratio $P_V = {{ V } \over {V+P}}$, where $V$ is the fraction 
of $\dst$ (vector) produced and $P$  the fraction of 
$\D$ (pseudoscalar), after all decays of heavier excited states, 
can be derived 
either from the ratio $f(\ch \to \dz) / f(\ch \to \dpl)$, or from
the $\dstpm$ and $\dpm$ rates. The latter leads to 
$$P_V=0.595\pm0.045.$$

This value is more consistent with the predictions of 
the thermodynamical approach~\cite{becattini}
and the string fragmentation approach of ref.~\cite{pei},
which both predict 0.66 for $P_V$, than from the naive expectation 
of 0.75 from counting the spin states. 

From the analysis of section 6, the
$P_V$ ratio is 
found to be $$0.60\pm0.19.$$
for $\dsp$ and $\dsstp$ mesons.

\boldmath
\subsection{Study of $\lc$ and $\dsp$ production}
\unboldmath

The c fractions in the $\dsp$ and $\lc$ samples are extracted by the same
method as for the $\dz$ and $\dpl$. The background fit includes the
Monte-Carlo shape and the proper normalisation of contributions from
$\dz$ and $\dpl$ to the invariant mass distributions.
The efficiency correction is similar, except that the 
data are counted in
8 $\xe$ bins between 0.2 and 1, allowing the fit of the $\xe$ distributions
to a Peterson-based shape for the purpose of extrapolation down to the
threshold. The breakdown of the systematic errors is very similar to
the one of the previous section and is given in the last two columns of
table \ref{t2}. As the relative $\bb$ to $\cc$ contributions are fitted
to the data, the uncertainty on the charm fraction is included in the
statistical error.  

The following $\dsp$ and $\lc$ product branching ratios in 
$\Zcc$ are obtained:
$$ \Rc \times f(\ch \to \ds) \times B(\dsp \to \phi \pi^+) \times
B( \phi \to \K^+ \K^- )  = 
 (0.352 \pm 0.057 \pm 0.021) \times 10^{-3} \\ $$
$$ \Rc \times f(\ch \to \lc) \times B(\lc \to \p \K \pi ) =
 (0.673 \pm 0.070 \pm 0.037 ) \times 10^{-3}. $$

Dividing these numbers by the Standard 
Model value of $\Rc$ and the branching fraction for the
reconstruction mode, 
the following fractions are obtained:
$$f(\ch \to \dsp) = 0.116 \pm 0.019 \pm 0.007 \pm 0.030 {\mathrm(BR)}$$
and 
$$f(\ch \to \lc) = 0.079 \pm 0.008 \pm 0.004 \pm 0.020 {\mathrm(BR)}.$$

Alternatively, one can use the sum of $\Rc \times f(\ch \to X_{\ch})$
over all the fundamental states of the various hadron species to
determine $\Rc$ as in section 8.

\boldmath
\section{Measurement of $\Rc$ from Charm Counting}
\unboldmath

By dividing the values of the product branching fractions found in
previous sections by the corresponding decay branching fraction,
 taken from \cite{pdg},
the following individual charm meson rates are obtained.

$$ \Rc \times f(\ch \to \dz)  = 
 0.0961 \pm 0.0029 \pm 0.0060 \pm 0.0023 (BR) \\ $$
$$\Rc \times f(\ch \to \dpl)  =
 0.0409 \pm 0.0013 \pm 0.0022 \pm 0.0033 (BR) \\ $$
$$\Rc \times f(\ch \to \dsp)  =
 0.0199 \pm 0.0032 \pm 0.0012 \pm 0.0050 (BR) \\ $$
$$\Rc \times f(\ch \to \lc)  =
 0.0135 \pm 0.0014 \pm 0.0007 \pm 0.0035 (BR) \\ $$

The small additional contribution from $\Xi_{\ch}$ and $\Omega_{\ch}$ is 
estimated to be 
$0.0034$ with a $50 \%$ uncertainty
using usual stangeness suppression factors.  
Summing all these contributions, a value of 
$$\Rc = 0.1738 \pm 0.0047 {\rm (stat.)}
\pm 0.0088 {\rm (syst.)} \pm 0.0075 {\rm (BR)}$$ 
is found where the systematic uncertainty is a linear sum of the
common contributions (nuclear interactions, $\dedx$, vertexing,
gluon splitting, charm fraction and fragmentation), added quadratically
to the other contributions. 
This is in good agreement with the Standard-Model 
expectation 0.1719~\cite{smrc} 
and the accuracy ($7\%$) of this measurement is 
similar to others~\cite{rcado,rc}.
The breakdown of the systematic errors on $\Rc$ is given in 
Table~\ref{trc}.

\begin{table}[htbp]
\caption{\label{trc}
{{\small Systematic errors on $\Rc$.}}}
\begin{center}
\begin{tabular}{|c|c|}
\hline
Source & $\delta \Rc$ \\
\hline
MC statistics   & $ 0.00207 $ \\
Charm physics   & $ 0.00176 $ \\
Fragmentation   & $ 0.00599 $ \\
Charm fraction  & $ 0.00096 $ \\
Gluon splitting & $ 0.00051 $ \\
$\dedx$         & $ 0.00156 $ \\
Nuclear interactions & $  0.00087 $ \\
Vertexing       & $ 0.00516 $ \\
Mass fits       & $ 0.00177 $ \\
\hline
$B(\dz \to \K^- \pi^+)$ & $ 0.0023$ \\
$B(\dpl \to \K^- \pi^+ \pi^+)$ & $ 0.0033$\\
$B(\dsp \to \phi \pi^+)$ & $ 0.0050$\\
$B(\lc \to \p \K^- \pi^+)$ & $ 0.0035$\\
baryons not decaying & \\
to $\lc$ & $ 0.0017 $ \\
\hline
Total internal & $ 0.0088$ \\
Total external & $ 0.0075$ \\
\hline
\end{tabular}
\end{center}
\end{table}

\section{Conclusions}
A significant improvement of the knowledge of charm production
in Z decays has been achieved. 
This is due, with respect to previous analyses, to a better
rejection of the combinatorial background and a better b/c separation, 
both coming from the use of the vertex detector.
The $\dstpm$ spectrum is 
measured and interpreted as a sum of three contributions: hadronisation
of a charm quark produced in a $\Zcc$ decay, decay of a bottom hadron
from a $\Zbb$ decay, and gluon splitting into a pair of heavy quarks
which hadronise or decay into a $\dstpm$.
This scheme is found to describe the data very well. 

The probability for a c quark to hadronise into a $\dstp$ meson
is found to be 
$$f(\ch \to \dstp)=0.233 \pm 0.010 \mathrm{(stat.)}
\pm 0.011 \mathrm{(syst.)}.$$

A fit to the spectrum yields the average fractional
energy of the $\dstpm$ in $\Zcc$ events
$${\langle \xe (\dst) \rangle }_{\cc} = 0.4878 \pm 0.0046 \pm 0.0061.$$

Using the heavy - light  hemisphere mass difference 
distribution the average number of gluon splitting events to 
charm quark pairs per Z hadronic event is found to be 
$$\bar{n}_{\g \to \cc} = (3.23 \pm 0.48 \pm 0.53)\%. $$ 
The production rates of the main charmed ground states have been 
measured. 
The effective ratio of Vector/(Vector+Pseudoscalar)
production rates in charmed mesons is found to be $0.595\pm0.045$,
confirming the rather low value found earlier~\cite{dspap}, but 
consistent with recent semi-phenomenological models.
For the charmed-strange mesons, a value of $P_V = 0.60 \pm 0.19$ is found.

Summing the contributions of all the fundamental charmed states, and
including a contribution from baryons not decaying to $\lc$, a measurement
of $\Rc$ is achieved:
$$\Rc = 0.1738 \pm 0.0047 {\rm (stat.)}
\pm 0.0088 {\rm (syst.)} \pm 0.0075 {\rm (BR)}.$$ This result, 
combined with other ALEPH measurements\cite{rc}, leads to 
$$\Rc = 0.1698 \pm 0.0069. $$

\vspace{1.5cm}
{\Large\noindent\bf\boldmath Acknowledgements}
\vspace{0.5cm}

We wish to thank our colleagues from the accelerator divisions for the
successful operation of the LEP machine, and the engineers and 
technical staff in all our institutions for their contribution to
the good performance of ALEPH. Those of us from non-member states
thank CERN for its hospitality.


%
%
\end{document}